\documentclass[sigconf,nonacm]{acmart} %

\usepackage{todonotes}
\usepackage{soul}
\usepackage{graphicx}
\usepackage{subcaption}
\usepackage{multirow}
\usepackage{algorithm}
\usepackage{algorithmic}
\usepackage{xspace}
\usepackage[inline]{enumitem}
\usepackage{url}

\newcommand{\approachName}{\textsc{GreenServ}\xspace}

\AtBeginDocument{%
  }

\acmConference[ICPE '26]{International Conference on Performance Engineering (ICPE)}{May 2026}{Florence, Italy}
\title{GreenServ: Energy-Efficient Context-Aware Dynamic Routing for Multi-Model LLM Inference}

\author{Thomas Ziller}
\affiliation{\institution{TU Wien}\city{Vienna}\country{Austria}}
\email{thomas.ziller@tuwien.ac.at}

\author{Shashikant Ilager}
\affiliation{\institution{University of Amsterdam}\city{Amsterdam}\country{The Netherlands}}
\email{s.s.ilager@uva.nl}

\author{Alessandro Tundo}
\affiliation{\institution{TU Wien}\city{Vienna}\country{Austria}}
\email{alessandro.tundo@tuwien.ac.at}

\author{Ezio Bartocci}
\affiliation{\institution{TU Wien}\city{Vienna}\country{Austria}}
\email{ezio.bartocci@tuwien.ac.at}

\author{Leonardo Mariani}
\affiliation{\institution{University of Milano-Bicocca}\city{Milan}\country{Italy}}
\email{leonardo.mariani@unimib.it}

\author{Ivona Brandic}
\affiliation{\institution{TU Wien}\city{Vienna}\country{Austria}}
\email{ivona.brandic@tuwien.ac.at}

\renewcommand{\shortauthors}{Ziller et al.}

\begin{abstract}

Large language models (LLMs) demonstrate remarkable capabilities, but their broad deployment is limited by significant computational resource demands, particularly energy consumption during inference. Static, one-model-fits-all inference strategies are often inefficient, as they do not exploit the diverse range of available models or adapt to varying query requirements.

This paper presents \approachName, a dynamic, context-aware routing framework that optimizes the trade-off between inference accuracy and energy efficiency. \approachName extracts lightweight contextual features from each query, including task type, semantic cluster, and text complexity, and routes queries to the most suitable model from a heterogeneous pool, based on observed accuracy and energy usage. We employ a multi-armed bandit approach to learn adaptive routing policies online. This approach operates under partial feedback, eliminates the need for extensive offline calibration, and streamlines the integration of new models into the inference pipeline.

We evaluated \approachName across five benchmark tasks and a pool of 16 contemporary open-access LLMs. Experimental results show that \approachName consistently outperforms static (single-model) and random baselines. In particular, compared to random routing, \approachName achieved a 22\% increase in accuracy while reducing cumulative energy consumption by 31\%. Finally, we evaluated \approachName with RouterBench, achieving an average accuracy of 71.7\% with a peak accuracy of 75.7\%. All artifacts are open-source and available here: 
\href{https://github.com/TZData1/llm-inference-router}{GitHub}

\end{abstract}


\begin{document}

\maketitle

\section{Introduction}\label{sec:introduction}

The rise of large language models (LLMs) is considered one of the major breakthroughs in machine learning, opening a new era of artificial intelligence (AI) with new capabilities such as human-like text and image generation. LLM-based autonomous agents have become an integral part of various applications and workflows, accelerating automation in areas such as customer service and market research~\cite{mckinsey2025stateofai}. However, training and the use of LLMs require a considerable amount of resources (e.g., energy) that raise several concerns about their computational sustainability~\cite{openai2024gpt4}.

Although approaches to reducing training costs for LLM have received a lot of attention, inference resource demands are often overlooked. In fact, the cumulative inference energy can exceed the training energy: a single ChatGPT query is estimated to consume approximately 2.9 Wh of energy for a total amount of 10 TeraWh per year \cite{ieaelectricity2024}.
Data centers hosting both training and inference already draw an appreciable share of global electricity, and their demand continues to grow~\cite{iea2025energyai}.

Current LLM inference often relies on static \emph{one-model-fits-all} strategies, routing all queries to the same large model regardless of complexity or quality~\cite{chenFrugalGPTHowUse2023}. While this dominates industry use, it wastes resources: studies show that many noncritical tasks (e.g., basic translation) can be handled by smaller, cheaper models with minimal quality loss~\cite{frantar2023optq, srivastava2022beyond}. The open source ecosystem adds to the challenge by offering over 230,000 text generation models on platforms like HuggingFace~\cite{huggingface}, including fine-tuned variants and optimized architectures (quantized, distilled, etc.), creating opportunities but complicating the decision process.

In fact, selecting an optimal LLM is not trivial. First, non-expert users often lack the technical expertise or explicit criteria needed to evaluate trade-offs between accuracy, cost, and latency, leading many to default to larger models assuming better capabilities. Second, the landscape is highly dynamic: leaderboards such as the HuggingFace Open LLM\footnote{\url{https://huggingface.co/open-llm-leaderboard}} and the CRFM HELM\footnote{\url{https://crfm.stanford.edu/helm/capabilities/latest/}} show that the top 50 models vary widely in size and specialization, and choices that are near-optimal today may become outdated within months. Third, performance is highly task-dependent; for example, smaller models can match or even outperform larger ones on focused tasks such as MMLU~\cite{hendrycks2021measuring}, but tend to underperform on broader challenges such as summarization~\cite{fu-etal-2024-tiny}.

To overcome the limitations of single model inference and improve the efficiency and performance of LLM inference, researchers have introduced two main computation paradigms: \emph{model cascading}~\cite{dohan2022languagemodelcascades} and
\emph{routing-based inference}~\cite{ong2024routellm}. Cascading approaches attempt to reduce cost by initially using a small, lightweight model to handle the request while then iteratively selecting more capable models until the output does not meet predefined quality thresholds. Although this method can improve efficiency, it often involves multiple inferences per request by design, significantly increasing both latency and computational cost~\cite{chenFrugalGPTHowUse2023}.

On the other hand, routing-based methods, such as RouteLLM~\cite{ong2024routellm}, MixLLM~\cite{wang2025mixllm}, LLMBandit~\cite{li2025llm}, and Eagle~\cite{zhao2024eagleefficienttrainingfreerouter}, aim to assign each inference request to the most appropriate model in a single step, based on a learned/heuristic decision process. Although these approaches show promising results, they are still affected by key limitations.

\textbf{Limited continual learning.} Most routers lack continual learning capabilities, operating statically after initial calibration, making them vulnerable to query distribution shifts and model degradation.

\textbf{Reliance on proxy cost metrics.} Quality-cost trade-offs often rely on synthetic proxies (API prices, token budgets) over actual production metrics like GPU or energy use, not directly measuring real resource consumption and limiting dynamic optimization.

\textbf{Underutilization of new models due to calibration overhead.} The availability of open-source model repositories with a diverse range of capabilities is constantly increasing, but many remain underexplored due to the calibration overhead associated with incorporating them. In fact, no approach supports zero-calibration integration, leading to underutilization of available model capabilities despite growing repository diversity.

To overcome these limitations, we propose \approachName, a \emph{dynamic, context-aware LLM inference routing framework} that assigns user queries to the most suitable model in its pool based on lightweight \emph{contextual features} (i.e., task type, semantic context, and text complexity) extracted from incoming user queries and learned knowledge about models' performance (i.e., accuracy and energy consumption). \approachName learns an \emph{adaptive routing policy online} to select the most suitable model at runtime using a \emph{contextual multi-armed bandit (MAB) algorithm}~\cite{LangfordZ07}. This enables online integration of new models without requiring extensive offline calibration.

Experimental results show that \approachName consistently outperforms single-model and random baselines by achieving superior energy-accuracy operating points. Compared to random routing, \approachName achieved accuracy gains of 22\% while reducing cumulative energy consumption by 31\%. Furthermore, results show that \approachName operates consistently close to or beyond the static optimal solutions, indicating effective control of the precision-energy trade-off based on the configurable parameter $\lambda$. The ablation study of contextual characteristics shows a substantial impact of the task type characteristic, dropping the median cumulative regret to $\approx 400$. This identifies the task type as the most informative component of the context to guide model selection in our setup. %
Moreover, the results confirm a successful adaptation of \approachName to the introduction of models into the model pool by integrating new and better-performing models into its routing strategy. Finally, the overhead results show that the total average overhead per query ranges between 6.68 and 7.77 ms, when processed sequentially, which indicates a negligible overhead for \approachName compared to the general inference time.
Finally, when evaluated with RouterBench~\cite{hu2024routerbench}, \approachName achieves an average AIQ and accuracy of 0.607 and 71.7\%
respectively, with a peak accuracy of 75.7\%.

In summary, this work provides the following key contributions.

\textbf{An adaptive context-aware LLM routing framework}. We propose an LLM routing framework capable of effectively balancing the trade-off between accuracy and energy consumption while meeting latency requirements. By leveraging a MAB algorithm, it assigns user queries to the most suitable available model, and it is capable of integrating new and better performing models into its routing strategy without requiring expensive offline calibration.

\textbf{A multi-feature query context representation}. We propose a multi-feature query representation (i.e., task type, semantic context, and text complexity) as a structured context vector, and study the impact of both single and combined features via ablation.

\textbf{Comprehensive baseline evaluation}. We employ LinUCB for model selection and evaluate \approachName's performance against multiple baselines including static routing strategies (random, smallest model, largest model, most accurate model) and alternative MAB approaches (\(\epsilon\)-Greedy, Contextual Thompson Sampling) using 5 benchmark tasks and a pool of 16 open-access LLMs from HuggingFace. In addition, we also performed a trade-off analysis to understand how \approachName behaves for different accuracy-energy-consumption ratio.

\textbf{An extensive empirical evaluation}. In addition to the ablation study for context characteristics, we performed specific experiments to study the adaptability to model addition, performed an overhead analysis, and evaluated \approachName using RouterBench~\cite{hu2024routerbench}. Finally, we analyzed the time and space complexity of our router agent.

The remainder of the paper is organized as follows. ~\S ~\ref{sec:related-work} discusses and categorizes related work, identifying key research gaps. ~\S ~\ref{sec:def-prelims} provides definitions and the necessary preliminary details for our contextual routing problem. ~\S ~\ref{sec:MOOP} formalizes the problem statements. ~\S ~\ref{sec:approach} presents \approachName, including its system architecture and a thorough description of the proposed solution. ~\S ~\ref{sec:implementation} outlines the implementation details. ~\S ~\ref{sec:evaluation} reports the empirical evaluation and results, discusses the main findings, and highlights current limitations. Finally, ~\S ~\ref{sec:conclusions} offers concluding remarks and directions for future work.

\section{Related Work}\label{sec:related-work}

Inference routing optimizes LLM  inference by assigning specific models to handle different types of queries~\cite{ong2024routellm}. The process maps input requests to particular LLMs from a heterogeneous pool, which can vary in terms of parameter sizes, architectures, or optimization levels. This approach ensures that computational resources are allocated according to the characteristics of each query.
\\
\textbf{Static Routing Systems.} Early LLM routing works used static, pre-deployment calibration strategies. For example, Tryage~\cite{hari2023tryage} used BERT embeddings for classification-based routing, achieving 50.9\% accuracy. TABI~\cite{yang2023tabi} focused on complexity-based routing to reduce latency by 21-40\%. RouterBench~\cite{hu2024routerbench} introduced an evaluation framework, while Hybrid LLM~\cite{ding2024hybrid} leveraged meta-learning to reduce large model calls by 40\%. Characterized by pre-deployment calibration (PD) and fixed policies (FP), these methods lack adaptability to dynamic environments and evolving models.
\\
\textbf{Embedding Representations and Learning-based Routing Systems.} Recent work on model routing focuses on advanced training methodologies. For example, RouteLLM \cite{ong2024routellm} introduced preference data-based routing using matrix factorization, demonstrating a cost reduction of up to 70\% while preserving performance. Smoothie~\cite{guha2024smoothie} introduced label-free routing via embedding-based similarity comparison and latent variable graphical models, eliminating the need for labeled data. RouterDC~ \cite{chen2024routerdc} employs dual contrastive learning for query-based routing, and EmbedLLM~\cite{zhuang2025embedllm} developed compact model representations using encoder-decoder architectures. Furthermore, GraphRouter~\cite{feng2025graphrouter} uses graph neural networks to model task-query-LLM interactions. These systems demonstrated improved routing accuracy through advanced representation learning. For new model additions, they lack adaptability for fast learning and the addition of models at runtime.
\\
\textbf{Dynamic and Adaptive Routing Systems.} The most recent works have focused on dynamic adaptation and runtime learning capabilities. TensorOpera \cite{tensoropera2024router} employed K-nearest neighbors for efficient inference, while Universal Model Routing \cite{jitkrittum2025universal} introduced cluster-based representation for unseen test-time LLMs, achieving dynamic model integration. LLMBandit \cite{li2025llm} proposed preference-conditioned dynamic routing using a multi-armed bandit formulation, enabling runtime preference specification and achieving significant cost reductions. MixLLM \cite{wang2025mixllm} presented the contextual-bandit-based routing, incorporating tag-enhanced embeddings and continual learning capabilities, 
achieving 97.25\% of GPT-4's quality at 24.18\% of the cost. 

 These systems represent significant progress in adaptive routing, moving from static configurations toward dynamic, learning-based approaches. Our approach extends this trajectory by combining minimal offline calibration with continuous online learning (hybrid calibration), enabling policy evolution through contextual bandits, supporting dynamic model integration without retraining, and optimizing multiple objectives using direct energy measurements rather than proxy metrics. We present a comprehensive routing framework capable of handling complex, evolving LLM ecosystems while maintaining near Pareto-optimal accuracy-energy trade-offs.

\section{Preliminaries and Problem Formulation}\label{sec:def-prelims}

In this section, we formalize the routing problem, which aims to dynamically assign incoming requests to a single model from a pool of heterogeneous LLMs while balancing two primary objectives: \emph{accuracy} and \emph{energy-consumption}. In the following subsections, we provide details about metrics and problem formulation.

\subsection{Metrics}\label{PM}   
The following section specifies how we define and measure accuracy and efficiency, which are then condensed into a MOOP through scalarization (see ~\S ~\ref{POS-OS}) to form the basis of our routing problem. Since the routing policy aims to balance the accuracy of the model's output against the energy consumed, we require metrics that quantify both and are measurable in an online setting. The following subsections detail these core metrics.

\subsubsection{Accuracy}
We denote \( Acc_m(q_t) \) as the \emph{accuracy} of model \( m \)'s response to query \( q_t \). Defining a single accuracy metric across all LLM tasks is challenging. For tasks with clear ground truth, objective metrics like Exact Match (EM)~\cite{rajpurkar2016squad}, ROUGE~\cite{lin2004rouge}, or BLEU~\cite{papineni2002bleu} are applicable. However, evaluating open-ended generation often requires subjective assessments (e.g., user feedback), which are difficult to automate and simulate reliably~\cite{chang2024survey}. Therefore, for deterministic evaluation within the scope of this work, we focus exclusively on tasks where accuracy can be measured objectively against an available ground truth using such unambiguous metrics. We assume \( Acc_m(q_t) \) is normalized to the interval \( [0,1]\), where higher values indicate a higher accuracy.

\subsubsection{Energy Consumption}
Efficiency in our setting refers to the effective utilization of computational resources to generate responses. It can be assessed through different aspects such as processing speed, memory footprint, or energy consumption. Among these factors, energy consumption stands out as a traceable and broadly relevant indicator for evaluating efficiency. Generally, the more efficient a system, the less energy consumption is required to produce responses of comparable quality, given the same hardware and software configuration.

Thus, energy consumption will serve as our proxy metric to judge efficiency. It can be formally defined by integrating the \emph{instantaneous power draw} \(P_m(t)\) of model \( m \) over inference duration \(T_{\text{proc}(m, q_t)}\):
\begin{equation}
    C_m(q_t)
    = \int_{0}^{T_{\text{proc}}(m,q_t)}
        P_m(\tau)\,d\tau
\end{equation}

This formulation provides the mathematical foundation for our multi-objective optimization. In practice, we measure \( C_m(q_t) \) directly via GPU power monitoring (§\ref{sec:implementation}).

\subsection{Problem Formulation}\label{sec:MOOP}
We consider an inference system that processes a sequential stream of user queries \(\{q_t\}_{t=1}^{T}\). Each query \( q_t \) arrives at a discrete timestep \( t = 1,2,\dots,T \), where T is the total number of queries.

Each query \( q_t \) may vary in specific characteristics (e.g., task type, text complexity), and thus, may require different levels of model capacity to be answered effectively. We assume access to a pool of \( K \) heterogeneous candidate LLMs:
$M = \{m_1, m_2, \dots, m_K\}$.
Each \(m_k \in M\) is a model with distinct characteristics. Possible characteristics include (i) architecture and parameter count, (ii) fine-tuning on domain (e.g., medical, legal), and (iii) quantization level (e.g., 8-bit, 4-bit precision), among others. In Equation \ref{eq:mainpf}, when query \( q_t \) arrives, the system's routing policy \( \pi \) selects exactly one model \( m_t \) for inference.

\begin{equation}\label{eq:mainpf}
    m_t = \pi(q_t).
\end{equation}

The policy $\pi$ aims to balance two key objectives -- accuracy and energy efficiency -- based on measured historical performance of each model in related contexts.

\subsubsection{Multi-Objective Optimization with Latency Constraints}\label{POS-OS}
Our routing problem balances accuracy and energy consumption as two competing objectives while respecting latency constraints that ensure acceptable Quality-of-Service (QoS)~\cite{zeng2004qosaware}.

Inference latency represents the time from query submission to response. However, the latency components, such as data transfer and queuing delays, depend on specific deployment environments and operating conditions, and introduce strong deviations. Therefore, we model the controllable latency components: optimization overhead \( L_{opt}(q_t) \) and inference processing time \(L_{m}(q_t)\):

\begin{equation}
    L_\text{total}(m, q_t) =  L_{opt}(q_t) + L_{m}(q_t).
\end{equation}

Users typically tolerate latency up to a threshold \( L_{max, t}\), beyond which satisfaction declines sharply~\cite{nah2004study}. We therefore define the set of feasible models for query \( q_t \) as:
\begin{equation}
    M_t^* = \{ m \in M | L_m(q_t) \le L_{max,t}\}.
\end{equation}

A model exceeding \( L_{max,t}\) is considered \emph{infeasible} and is discarded in the candidate selection at time step \( t \), analogous to established QoS techniques ~\cite{zeng2004qosaware}.

To handle the accuracy-energy trade-off, we apply the \emph{Weighted Sum Method}~\cite{marler2004survey}. For our routing problem, we combine accuracy and energy consumption for a query \(q_t\) as follows:
\begin{equation}\label{eq:reward}
    r_t(m,q_t) = \alpha Acc_m(q_t) - \beta C_m(q_t), \;\text{subject to}\; m \in M_t^*. 
\end{equation}

where \(\alpha = 1 - \lambda, \; \beta = \lambda,\; 0 \leq \lambda \leq 1\). This parameter \(\lambda\) conveniently allows interpolation between accuracy-only (\(\lambda=0\)) and energy-
only (\(\lambda=1\)) policies. While this approach assumes a fixed rate of trade-off between objectives, it enables efficient, adaptive decision-making in practice. In our experiments, we perform a parameter sweep over \(\lambda\) to evaluate these trade-offs. However, solving Equation~\ref{eq:reward} online requires complete observations across models, hardware settings, and tasks; an assumption rarely feasible in practice. This necessitates a learning-based routing strategy under partial feedback. Consequently, in the next section, we describe how we tackled this challenge using a contextual multi-armed bandit framework.

\subsubsection{Bandit Problem Formulation}\label{CBF}

 In dynamic online settings, we need to learn optimal routing policies without exhaustive prior knowledge, i.e., we can only observe the performance of the selected model \( m_t \) on each query \( q_t \) as outcomes for unselected models remain unknown. This \emph{partial feedback} structure~\cite{lattimore2020bandit} strongly motivates the application of Multi-Armed Bandit (MAB) algorithms. Each query \( q_t \) corresponds to a decision point, and each feasible model \( m \in M^*_t \) represents an arm. After selecting \( m_t \), the system observes the scalarized reward \( r_t \) based on accuracy and energy (Eq.~\ref{eq:reward}).
Moreover, we provide a key extension of classical MAB algorithms, with the inclusion of context features (described in~\S~\ref{sec:methodology_qcg}). Instead of treating all queries identically, \emph{contextual bandits} leverage relationships between input characteristics and reward outcomes~\cite{li2010contextual}. Utilizing context vector \( x_t \) extracted from \( q_t \), we learn a policy \( \pi: x_t \rightarrow m_t \) that selects a feasible model \( m_t \in M^*_t \). Since different queries may demand varying model capacities, this approach is expected to outperform static strategies. Over time, the policy continuously gains information and refines its estimates to approximate the true reward functions for \( x_t \), allowing the system to adapt to new query distributions and handle integration of new models in online settings.

To evaluate the performance of a policy, we measure the performance gap using the concept of \emph{regret}~\cite{lattimore2020bandit}. The oracle policy with complete knowledge of model performances would have selected the optimal model:

\begin{equation}
    m_t^* = \arg\max_{m \in M_t^*} r_t(m, q_t),
\end{equation}

where \( r_t(m_t, q_t) \) is the reward function introduced in~\S~\ref{POS-OS} and \( M_t^* \) defines the set of feasible models at time step \( t \).  Thus, \emph{instantaneous regret} is defined as:

\begin{equation}
  \Delta_t = r_t(m_t^*) - r_t(m_t),
\end{equation}

where \( m_t \) is the model selected by the routing policy. If the optimal model was chosen, regret is zero. After \( T \) iterations, cumulative regret is defined as:

\begin{equation}
    \text{Regret}(T) = \sum_{t=1}^{T} \big( r_t(m_t^*) - r_t(m_t) \big).
\end{equation}

The routing strategy aims to minimize \( \text{Regret}(T) \). Since MAB is optimizing a multi-objective trade-off between accuracy and energy efficiency, minimizing regret aligns with approximating context-specific Pareto fronts and discarding consistently dominated models.

\section{\approachName: Learning Energy-Efficient Context-Aware Dynamic Routing}\label{sec:approach}
In this section, we first define the system model along with its core components. Subsequently, we present the solution methodology for \approachName.

\subsection{System Model}\label{section:system_overview}

\begin{figure}[!ht]
    \centering
    \includegraphics[scale=0.65]{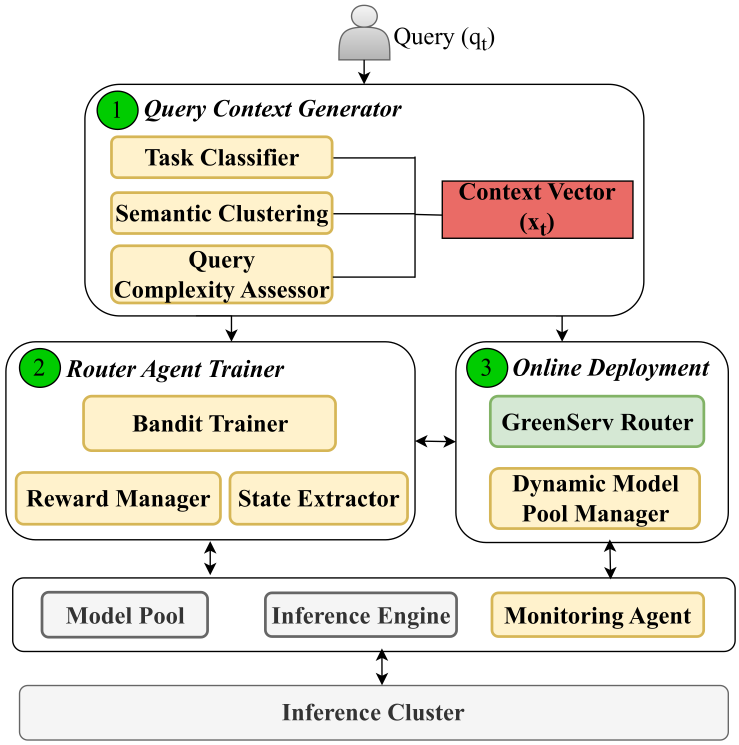}
    \caption{The system model of \approachName.}
    \label{fig:system-architecture}
\end{figure}

Figure~\ref{fig:system-architecture} illustrates the high-level view of the \approachName's system model.  \approachName comprises three main components: (i) \texttt{Query Context Generator}, which extracts the necessary metadata to construct a query-specific context vector, capturing the unique characteristics of each query; (ii) \texttt{Router Agent Trainer}, where trains an agent using bandit learning to identify efficient routing configurations and (iii) \texttt{Online Deployment}, in which the trained router is deployed to handle inference requests in real time. It is important to note that \approachName is capable of adapting to the addition of new models to the existing model pool.

\subsection{Query Context Generator}\label{sec:methodology_qcg}  

The \texttt{Query Context Generator} processes each query through three components: \emph{Task Classifier}, \emph{Semantic Clustering}, and \emph{Query Complexity Assessor}. These modules extract three features: \emph{Task Type} (query intent~\cite{hu2024routerbench}); \emph{Cluster} (semantic context via embedding-based clustering); 
and \emph{Complexity Score} (textual complexity).
These are then combined into context vector $x_t$.
By encapsulating multiple dimensions of the query,
 $x_t$ allows the router to make informed decisions about model selection to effectively balance inference performance against energy efficiency.

\subsubsection{Task Classifier}  We rely on a lightweight text classification approach to identify high-level task types (e.g., summarization, QA). Specifically, we train a \emph{Logistic Regression} (LR) model on top of semantic embeddings~\cite{bishop2006pattern}.

We extract the instruction text \( q_{\text{instr},t} \) from the initial lines of the prompt \( q_t \), following the common structure of instruction-based tasks. Its embedding \( e_{\text{instr},t} = \operatorname{embedding}(q_{\text{instr},t}) \) is computed using a pre-trained transformer model~\cite{reimers-gurevych-2019-sentence}, yielding a semantic vector representation. 

An LR model is then trained using cross-entropy loss over labeled pairs \( (e_{\text{instr},t}, l_t) \), where \( l_t \) denotes the ground-truth label~\cite{bishop2006pattern}. The probability distribution is modeled as \(  p(l \mid e_{\text{instr},t}) = \sigma (W e_{\text{instr},t} + b) \) with parameters \( W \) and \( b \) optimized to learn a decision boundary for classification. 

For training data, we sample a small portion of our evaluation dataset described in ~\S \ref{EDDS}. Each query is annotated with a ground-truth task label based on its dataset of origin. We split the data into training and validation subsets, train the LR model, and evaluate its performance on the validation set using standard classification metrics such as the F1 score~\cite{power2011fmeasure}. After training, we store the LR parameters for further use. The resulting model is a simple classifier that extracts the discrete task label. We specifically choose LR for its computational simplicity, speed, and low resource footprint to avoid excessive offline overhead.

\subsubsection{Semantic Clustering}
     For grouping the queries into domain clusters, we begin with embedding the entire query \( q_t \) using a pre-trained transformer-based embedding model~\cite{reimers-gurevych-2019-sentence} as \( e_{full,t} =  embedding (q_t) \).
     Utilizing these full embeddings, we perform an online K-Means clustering algorithm~\cite{bottou1994convergence} with a fixed number of clusters, \(K\) (specified in ~\S  \ref{sec:experimental_setup}), to group queries based on semantic similarity. The algorithm assigns a query to the cluster with the most similar centroid by maximizing the cosine similarity between the query's embedding and the cluster centroids:
     
     \begin{equation}
        c_t = \arg\max_c \frac{e_{\text{full}, t} \cdot \mu_c}{\|e_{\text{full}, t}\| \|\mu_c\|}.
     \end{equation}
     
     The centroids \( \mu_c \) are updated online based on the number of points assigned to the cluster so far (\(N_c\)) as new queries \( e_{\text{full},t} \) are observed:
     
     \begin{equation}
     \mu_{c_t} \leftarrow \mu_{c_t} + \frac{1}{N_{c_t}+1} (e_{\text{full},t} - \mu_{c_t}),         
     \end{equation}

     where \( c_t \) is the cluster assigned to the current query \( e_{\text{full},t} \) and \( N_{c_t} \) is the count of previous points assigned to that cluster. This implements a standard incremental update with a decaying learning rate.

     Online K-Means is chosen for its ability to adapt centroids incrementally without storing all past embeddings or requiring pre-determined clusters. The number of clusters K is fixed (value specified in ~\S \ref{sec:experimental_setup}) as a tunable hyperparameter to balance granularity and stability. Initial centroids are chosen from the first K distinct query embeddings. The online update allows for the cluster centroids to incrementally adapt to shifts in the topics of incoming queries over time. Cluster label \( c_t \) represents the semantic domain (or potentially topic) of a query \( q_t \).

 \subsubsection{Query Complexity Assessor}
     We calculate a single numeric score that represents the complexity of query \( q_t \) based on the Flesch Reading Ease formula~\cite{flesch1948new}:
     
     \begin{equation}
         p(q_t) = 206.835 - 1.015 \cdot \left(\frac{\text{Words}_t}{\text{Sentences}_t}\right) - 84.6 \cdot \left(\frac{\text{Syllables}_t}{\text{Words}_t}\right).
     \end{equation}

     For each \( q \), this results in a value \( \in [0, 100] \) where a low score indicates high text complexity. To convert this numerical score into a categorical feature suitable for the context vector, we bin the scores into \(N_{\text{bins}}\) distinct categories using equal-width binning. Again, the specific number of bins (\(N_{\text{bins}}\)) and the corresponding score ranges are detailed in the evaluation setup (~\S \ref{sec:experimental_setup}).

    \subsubsection{Context Vector}
    Our context vector encodes relevant characteristics of a query as \( x_t = [l_t, c_t, p_t] \). The resulting vector is forwarded to the router (described in ~\S ~\ref{sec:router_agent}), which aims to leverage these features to make more informed, query-dependent model choices. For the contextual bandit algorithms introduced in the next section, categorical features \(l_t\), \(c_t\), and \(p_t\) are converted into a numerical feature vector using one-hot encoding~\cite{bishop2006pattern}. Additionally, we add an intercept term (bias) by appending a constant value of 1, and thus, the resulting context vector \( x_t \in \mathbb{R}^d \) has dimension \( d = N_{\text{tasks}} + K + N_{\text{bins}} + 1 \). The specific parameter values used in our experiments were determined by non-exhaustive tuning described in ~\S ~\ref{sec:hyperparameter-tuning} and are intended to balance feature granularity with dimensionality. The specific number of task types (\(N_{\text{tasks}}\)) is determined by the evaluation datasets defined in ~\S ~\ref{EDDS}.

\subsection{Router Agent Trainer}\label{sec:router_agent}

To handle unseen queries and out-of-domain distributions, we adopt a multi-armed bandit (MAB) approach to learn a policy for selecting models from the feasible set \( m_t \in M_t^* \). We employ LinUCB~\cite{li2010contextual}, treating each model \( m \in M \) as an arm and selecting model \( m_t \) based on context vector \( x_t \).

\noindent \textbf{State Extractor.} \approachName constructs the state using three parameters: accuracy \(Acc_m(q_t)\), energy consumption \(Em(q_t)\), and latency \(L_m(q_t)\). Accuracy and energy are captured by interacting with the inference engine and monitoring agent. For latency, we use the predefined maximum output tokens (\(\text{MaxNewTokens}\)) for the query's task type as a conservative estimate.

\noindent \textbf{Reward Manager.} The MAB maximizes the scalarized reward (Equation~\eqref{eq:reward}) parameterized by $\lambda \in [0,1]$, where $\alpha = 1 - \lambda$ and $\beta = \lambda$. This balances accuracy ($\lambda = 0$) and energy efficiency ($\lambda = 1$). Performance is evaluated by cumulative regret over \( T \) steps:

\begin{equation}
  R(T) = \sum_{t=1}^{T} [r_t(m_t^*, q_t, \lambda) - r_t(m_t, q_t, \lambda)],  
\end{equation}

where \( m_t = \pi(x_t) \) is the chosen model and the optimal choice at time \( t \) is \( m_t^* = \arg\max_{m \in M_t^*} r_t(m, q_t, \lambda) \).

 \noindent \textbf{Bandit Trainer.}
\begin{algorithm}[!ht]
\caption{GreenServ: Context-Aware Routing for Multi-Model LLM Inference}
\label{alg:greenserv}
\begin{algorithmic}[1]
\REQUIRE Query stream $\{q_t\}_{t=1}^T$, model pool $\mathcal{M}$, LinUCB algorithm $\mathcal{A}$, parameters $\lambda$
\ENSURE Model selections $\{m_t\}_{t=1}^T$
\STATE Initialize LinUCB parameters $\{\mathbf{A}_m, \mathbf{b}_m\}$ for each $m \in \mathcal{M}$
\STATE Initialize task classifier $W$, cluster centroids $\boldsymbol{\mu}$
\FOR{$t = 1$ to $T$}
    \STATE $\mathbf{x}_t \gets$ GenerateContext$(q_t)$ \hfill \COMMENT{Task, cluster, complexity}
    \STATE $m_t \gets$ SelectModel$(\mathbf{x}_t, \mathcal{M}_t^*, \mathcal{A})$ \hfill \COMMENT{LinUCB routing}
    \STATE response $\gets$ InferenceExecution$(m_t, q_t)$
    \STATE accuracy, energy, latency $\gets$ Monitor(response) \hfill \COMMENT{Performance metrics}
    \STATE $r_t \gets (1-\lambda) \cdot \text{accuracy} - \lambda \cdot \text{energy}$
    \STATE UpdateMAB$(\mathbf{A}_{m_t}, \mathbf{b}_{m_t}, \mathbf{x}_t, r_t)$
\ENDFOR
\RETURN $\{m_t\}_{t=1}^T$
\end{algorithmic}
\end{algorithm}

Algorithm~\ref{alg:greenserv} presents \approachName's context-aware routing. It takes query stream $\{q_t\}_{t=1}^T$, model pool $\mathcal{M}$, and trade-off parameter $\lambda$ as inputs. For each query, it extracts context vector $\mathbf{x}_t$, selects model $m_t$ using LinUCB, executes inference, computes reward $r_t$ (balancing accuracy and energy via $\lambda$), and updates MAB parameters for continuous online learning.

\approachName employs LinUCB~\cite{li2010contextual}, a contextual bandit algorithm that assumes a linear relationship between context and reward, $\hat{r}_m(\mathbf{x}_t) = \boldsymbol{\theta}_m^T \mathbf{x}_t$, and maintains parameters $\mathbf{A}_m \in \mathbb{R}^{d \times d}$ and $\mathbf{b}_m \in \mathbb{R}^d$ for each model. Parameters are estimated as $\hat{\boldsymbol{\theta}}_m = \mathbf{A}_m^{-1} \mathbf{b}_m$ and updated as $\mathbf{A}_{m_t} \gets \mathbf{A}_{m_t} + \mathbf{x}_t \mathbf{x}_t^T$ and $\mathbf{b}_{m_t} \gets \mathbf{b}_{m_t} + r_t \mathbf{x}_t$ after observing rewards.

LinUCB employs systematic uncertainty quantification for exploration. It augments the expected reward with an exploration bonus proportional to parameter uncertainty:
\begin{equation}
    m_t = \arg\max_{m \in \mathcal{M}_t^*} \left( \hat{\boldsymbol{\theta}}_m^T \mathbf{x}_t + \alpha \sqrt{\mathbf{x}_t^T \mathbf{A}_m^{-1} \mathbf{x}_t} \right),
\end{equation}

where $\mathbf{x}_t^T \mathbf{A}_m^{-1} \mathbf{x}_t$ quantifies the reward estimate variance in context $\mathbf{x}_t$. The upper confidence bound approach ensures exploration targets regions of high uncertainty over uniform randomness. For baseline comparison, we also implement $\epsilon$-Greedy~\cite{sutton1998reinforcement}, which employs random exploration with probability $\epsilon$ and greedy exploitation otherwise. Contextual Thompson Sampling~\cite{agrawal2013thompson} uses Bayesian posterior sampling over model parameters. Both baselines rely on the same linear reward model as LinUCB.

\textbf{Complexity Analysis.} Processing $T$ queries, \approachName incurs a time complexity of $O(T \cdot (l + |M|d^3))$, where $l$ is the input text length, $|M|$ the number of candidate models, and $d$ the feature vector dimension. 
Space complexity for LinUCB is $O(|M|d^2)$ to maintain parameter matrices for each arm. 
We provide a detailed analysis of time and space complexity in Appendix~\ref{appendix:complexity-analysis}.

\subsection{Online Deployment}\label{sec:online_deployment}
During online deployment, the \approachName router processes the query \( q_t \). It computes the context vector \( x_t \) for the given query and utilizes a trained Multi-Armed Bandit (MAB) agent to select a suitable model. Specifically, if the chosen model \( m_t \) is not already present in memory, the router interacts with the inference engine to load the model into GPU memory. Once the model is loaded, the inference engine generates a response for the query.

Note that multiple models can reside in memory simultaneously, and further optimizations may be applied to reduce the cost of model loading. However, these optimizations are beyond the scope of this work, as our primary focus is on model selection.

The system tracks energy consumption and latency from the beginning to the end of the inference process. In addition, it logs key metadata, including the number of input tokens and generated output tokens.
Moreover, if a new model is added to the model pool, \approachName can dynamically learn and adapt to the newly introduced model through online interaction with the agent trainer.

\section{Implementation}\label{sec:implementation}

We implement the \approachName prototype in Python~3.10. User requests are handled via an HTTP API built with \texttt{FastAPI}~\cite{fastapi}, queued for inference over an asynchronous connection to \texttt{Redis}~\cite{redis}, and stored in a \texttt{PostgreSQL}~\cite{postgres} database, which holds prompts, model identifiers, results, and metrics. To isolate performance characteristics, we process each request independently (\texttt{batch\_size = 1}) and assume all model weights are locally available at runtime.

Datasets are loaded through the HuggingFace \texttt{datasets} library ~\cite{huggingface}, with random sampling conducted via Python’s \texttt{random} package. For feature extraction, we compute transformer-based sentence embeddings using \texttt{sentence-transformers} ~\cite{sbert} (specifically, the \texttt{all-MiniLM-L6-v2} model). Basic classification is performed using Logistic Regression~\cite{bishop2006pattern}, and similar queries are clustered online via K-Means~\cite{bottou1994convergence}, both implemented with \texttt{scikit-learn} ~\cite{scikitlearn}. Text complexity is measured using \texttt{textstat} ~\cite{textstat} and the Flesch Reading Ease score~\cite{flesch1948new}, which we discretize via equal-width binning based on empirical score ranges.

We implement LinUCB~\cite{li2010contextual} as the core routing algorithm for \approachName, along with \(\epsilon\)-Greedy~\cite{sutton1998reinforcement} and CTS~\cite{agrawal2013thompson} as baseline alternatives for comparison. All strategies are implemented as custom Python classes, with internal operations performed on \texttt{NumPy} arrays ~\cite{numpy}. LLMs are stored locally and loaded as HuggingFace-compatible models using the \texttt{transformers} ~\cite{transformersdoc} library in conjunction with \texttt{PyTorch} ~\cite{pytorch}, using \texttt{bfloat16} precision~\cite{kalamkar2019study} to reduce GPU memory usage and ensure efficient inference. All models undergo a single warm-up inference post-load to account for lazy initialization that might otherwise skew latency measurements.

Prior work often relies on proxy metrics (API costs, token budgets). We measure actual GPU power draw in watt-hours for direct optimization of actual resource consumption using the \texttt{zeus} library~\cite{zeusgithub}. Inference latency is measured with Python's \texttt{time} module, excluding queueing, feature extraction, routing, and model loading. Accuracy is evaluated using Exact Match (EM) and ROUGE~\cite{lin2004rouge} via HuggingFace's \texttt{evaluate} library~\cite{evaluatedoc}. The code and associated artifacts are provided here: \texttt{GitHub repository}\footnote{\url{https://github.com/TZData1/llm-inference-router}}.

\section{Empirical Evaluation}\label{sec:evaluation}

This section presents our empirical evaluation. In particular, we first detail our experimental setup in Section~\ref{sec:experimental_setup} describing the testbed, the evaluation datasets and the LLMs we used in our model pool, the evaluation metrics, how we tune the hyperparameters, and the baselines we used for comparison. Further, we outline our experimental plan in ~\S ~\ref{sec:experimental_plan} and we present empirical results in ~\S ~\ref{sec:results_analysis}. Finally, ~\S ~\ref{sec:discussion} discusses the key findings, their implications, and the current limitations of the study.

\subsection{Experimental Setup}\label{sec:experimental_setup}

\subsubsection{Testbed}\label{sec:testbed}
We run our experiments on a server running Ubuntu 22.04.5 LTS and equipped with 512 GB of RAM, an NVIDIA A100 GPU with 80 GB VRAM (CUDA 12.2), and an AMD EPYC 9354P processor with 32 cores. The GPU supports compute optimized bfloat16 acceleration~\cite{kalamkar2019study}. 

\subsubsection{Datasets}\label{EDDS}
We selected five publicly available datasets encompassing a broad range of query types, complexity levels, and domains. For each dataset, 500 instances were uniformly sampled from the test set partition using a fixed random seed to ensure reproducibility. Specifically, we utilize the MMLU~\cite{hendrycks2021measuring} dataset for question answering, HellaSwag~\cite{zellers2019hellaswag} for situation completion, Winogrande~\cite{winogrande2025} for commonsense reasoning, GSM8K~\cite{cobbe2021training} for mathematical reasoning, and CNN / Daily Mail~\cite{hermann2015teaching} for summarization. Evaluation for MMLU, HellaSwag, Winogrande, and GSM8K is performed using the exact match metric, whereas the CNN / Daily Mail dataset is assessed with the ROUGE metric~\cite{lin2004rouge}.

\subsubsection{Model Pool}

We compose a pool of 16 publicly available LLMs, representing a range of parameter scales and model families.
Our model pool selection is guided by four criteria:

\begin{enumerate}
    \item Diversity in parameter counts, spanning from 0.5B to 34B, based on compatibility with our computational resources;
    \item Popularity of model families provided by leading vendors, namely Phi ~\cite{mpimodels}, Gemma ~\cite{gemmamodels}, Mistral ~\cite{mistralmodels}, Llama ~\cite{llamamodels}, Qwen ~\cite{qwenmodels};
    \item Availability of model weights for local deployment;
    \item Recency of publication to ensure state-of-the-art performance.
\end{enumerate}

The final model pool consists of five Qwen 2.5 models (0.5B, 1.5B, 3B, 7B, 14B), Mistral v0.3 7B, four Gemma 3 models (1B, 4B, 12B, 27B), two Llama 3.1 models (1B and 8B) and Llama 3.2 3B, two Phi models (4-mini 4B and 4 14B), as well as Yi 34B. Appendix~\ref{appendix:model-pool} provides a summary table of the LLMs evaluated in our experiments, grouped by model family, along with their parameter counts and Hugging Face identifiers (HF Handles).

\subsubsection{Evaluation Metrics}
We report \emph{mean normalized accuracy}, \emph{total energy consumption (Wh)}, \emph{model selection frequency}, and \emph{cumulative} and \emph{moving average regret} values in our experimental results. To estimate system overhead, we report \emph{mean latency (ms)} and \emph{model selection time}. Results include 95\% confidence intervals, where appropriate, to account for the inherent variance in LLM outputs and the MAB learning processes.

The \emph{normalized accuracy} is calculated through min-max normalization, which converts observed accuracy values to a range of $[0,1]$. This allows a consistent comparison across the different evaluation metrics employed in the selected datasets.

\begin{equation}
    \text{Normalized Accuracy} = \frac{Acc - Acc_{\text{min}}}{Acc_{\text{max}} - Acc_{\text{min}}}
\end{equation}

where \(Acc_{\text{min}}\) and \(Acc_{\text{max}}\) are determined from baseline profiling runs of representative models in our pool on the validation set for each specific task type. For establishing these bounds, we strategically selected models that are likely to represent accuracy extremes, i.e.,  using smaller, older models (e.g., Phi2-3B) to estimate minimal accuracy values and larger, newer models (e.g., Qwen2.5-32B) to estimate maximum accuracy thresholds.

\subsubsection{Hyperparameter Tuning}\label{sec:hyperparameter-tuning}
Prior to the main experiments, we conducted preliminary experiments to tune hyperparameters for LinUCB and baseline MAB algorithms, as well as feature extraction parameters ($K_{\text{cluster}}$, $N_{\text{bins}}$), evaluating their impact on cumulative regret. For LinUCB: $\alpha=0.1$ and $\lambda_{\text{reg}}=0.05$. For $\epsilon$-Greedy: $\epsilon_0 = 1.0$, $\delta = 0.98$, and $\epsilon_{\text{min}} = 0.01$. For CTS: $\sigma = 0.01$.

For feature extraction, we used $K=3$ semantic clusters and $N_{\text{bins}}=3$ text complexity bins, unless specified differently. After one-hot encoding and the addition of an intercept term as described in ~\S ~\ref{sec:approach}, this yields a context vector dimension of $d=12$ for the contextual bandit algorithms.

\subsubsection{Baselines}\label{sec:baselines}
We compare \approachName against the following four baselines.

\begin{enumerate}
    \item \emph{Random}. Random selection serves as an estimate of how the whole model population achieves accuracy and efficiency on average by randomly picking a model from the model pool for each query.
    \item \emph{Largest (Yi-34B)}. In many deployments, one finds the tendency to assume that larger models yield better results. This approach disregards resource demands and simulates real-world scenarios that place model accuracy above all else.
    \item \emph{Smallest (Qwen2.5-0.5B)}.
    A baseline with the smallest model shows a scenario where energy use and hardware requirements determine model choice. It indicates the accuracy loss associated when minimizing resource consumption.
  \item \emph{Highest Accuracy (Gemma-3-27B)}: This approach selects the model that achieves the highest possible average accuracy on benchmark tasks, without considering efficiency. It is derived through exhaustive profiling of all models in the model pool, and serves as the upper bound of achievable accuracy in a single-model setting.
 
 \item \emph{\(\epsilon\)-Greedy and Thompson Sampling}: These two serve as \approachName variants with two different MAB strategies. Although \approachName defaults to LinUCB, it allows extensibility to other bandit algorithms.  \(\epsilon\)-Greedy is a core exploration-exploitation method~\cite{sutton1998reinforcement}, with probability of \(\epsilon\), a random model is chosen (exploration), otherwise the model with the highest expected reward (exploitation). Thompson Sampling instead maintains a posterior over model performance~\cite{agrawal2013thompson}. It selects the model with the highest sampled reward from this distribution, balancing exploration and exploitation via \emph{Bayesian inference}. 

\end{enumerate}

\subsection{Experimental Plan}\label{sec:experimental_plan}
We study the effectiveness and efficiency of \approachName through five experiments.

\subsubsection{\approachName vs. Baselines}
This experiment studies the effectiveness of \approachName by comparing it against baselines presented in ~\S ~\ref{sec:baselines}. We investigate learning behavior based on cumulative and moving-average regret, and evaluate the resulting regret, mean normalized accuracy, and total energy consumption.%

\subsubsection{Trade-off Analysis (\(\lambda\) Sweep)}
This experiment studies how \approachName handles the trade-offs at different configurations by varying the \( \lambda \) parameter. We vary the \(\lambda\) parameter from 0 (accuracy only) to 1 (efficiency only) in increments of 0.1. For each value, we executed 20 runs for \approachName and each baseline MAB algorithm with all contextual features activated. %

\subsubsection{Impact of Contextual Features}
This experiment studies the impact of different contextual features on routing decisions, as not all features are expected to contribute equally to potential accuracy and efficiency gains. We run experiments with different feature configurations. In particular, context-free routing uses no features and the router learns only their global average reward. Single-feature routing only has information about one of the three context dimensions (i.e., task type, semantic cluster, text complexity) during the whole experiment run. Full-context routing leverages all features by using all derived query characteristics. We executed 50 runs for each configuration by employing LinUCB.

\subsubsection{Adaptability: Model Addition}
This experiment examines how \approachName effectively adapts to changes in the model pool to simulate real-world scenarios of regular model releases. We introduce a new model (Gemma-3-12b), which showed high reward scores in a previous experiment after 1000 queries, and investigate if and how extensively the system incorporates it. In this experiment, we use LinUCB with full features and \(\lambda=0.2\) to favor the new high-accuracy model.

\subsubsection{Overhead Analysis}
This experiment evaluates the overhead the routing mechanism itself introduces to determine whether the benefits outweigh the costs. In particular, we account for the average latency introduced by each step involved in the feature extraction and the routing decision process per query.

For evaluation, we use our evaluation dataset consisting of 500 samples from each of the five benchmarks (i.e., total sequence length of \(T=2,500\) queries per experiment run). Unless otherwise specified, each experiment runs for the full sequence length of \(T=2,500\).%

\subsection{Results}\label{sec:results_analysis}

\subsubsection{\approachName vs. Baselines}\label{sec:results_bandits}

\begin{figure}[!ht]
    \centering
    \begin{subfigure}[b]{0.95\linewidth}
        \centering
        \includegraphics[width=\linewidth]{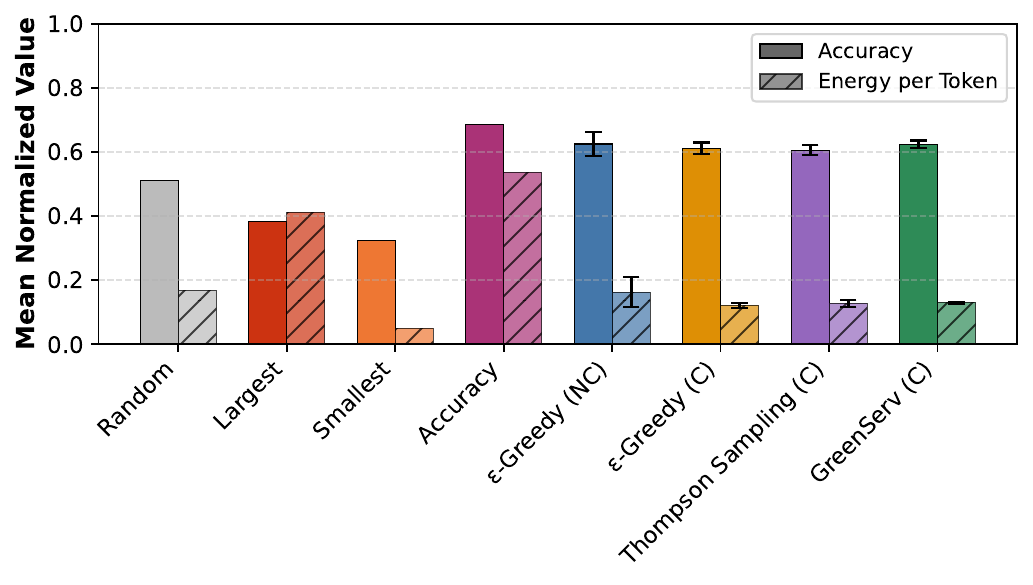} 
        \caption{}
        \label{fig:a5_bar_plot}
    \end{subfigure}
    \hfill
    \begin{subfigure}[b]{\linewidth}
        \centering
        \includegraphics[width=\linewidth]{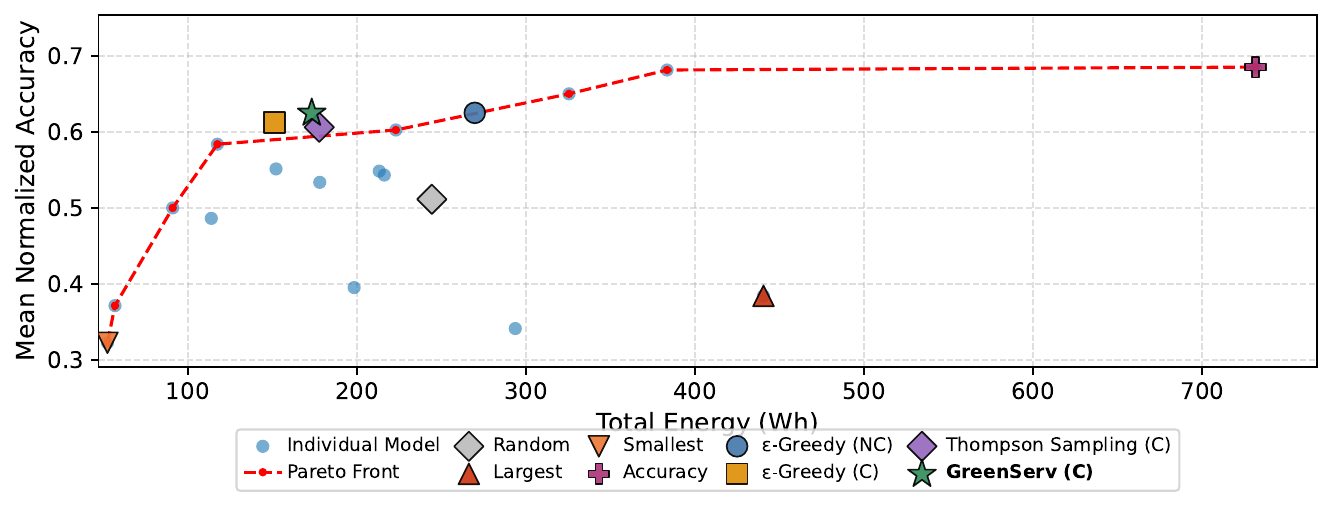} 
        \caption{}
    \label{fig:a5_pareto}
    \end{subfigure}
    \caption{Comparison of mean normalized accuracy (higher is better) and total energy consumption. \approachName and contextual baselines use all available features. Error bars represent 95\% confidence intervals. The static Pareto frontier in Figure~\ref{fig:a5_pareto} is shown for reference.}
    \label{fig:baseline-comparison}
\end{figure}

Figure~\ref{fig:a5_bar_plot} shows the mean normalized accuracy and energy consumption achieved by \approachName (\(\lambda=0.4\)) compared to baselines. The data are the results of 50 experiment runs and includes 95\% confidence intervals.  \approachName and contextual baseline algorithms consistently achieve higher accuracy at lower energy consumption when compared to static and non-contextual baselines. In particular,  \approachName achieves an accuracy of $\approx 0.65$ with significantly lower energy consumption ($\approx 165$ Wh). More specifically, \approachName and contextual baselines outperform the non-contextual \(\epsilon\)-Greedy by reaching higher accuracy (0.64-0.65 vs. 0.59), while reducing energy consumption by 29-38\%. Compared to the static baselines, improvements become even more prevalent with \approachName reducing energy consumption substantially compared to the random (31\%), largest (64\%), and accuracy (77\%) baselines, while simultaneously achieving superior accuracy to the random (\( \approx 0.51 \)), largest (\( \approx 0.39 \)), and smallest (\( \approx 0.33 \)) baselines. The confidence intervals for both accuracy and energy consumption largely overlap across \approachName and the contextual baselines, indicating comparable performance and validating our selection of LinUCB based on its superior accuracy on external validation.

Similarly, Figure~\ref{fig:a5_pareto} illustrates the trade-off of the same metrics but for a single run. The static Pareto front (red dashed line) is shown for reference. %
Predominantly, \approachName (LinUCB) and the contextual baselines (Contextual \(\epsilon\)-Greedy and Contextual Thompson Sampling) position themselves closely together in the more optimal top-left region. %
\approachName and the contextual baselines surpass the static Pareto front, which demonstrates that using context for dynamic routing can allow superior accuracy-efficiency balance by effectively combining the usage of multiple models.

\begin{figure}
    \centering
    \includegraphics[width=\linewidth]{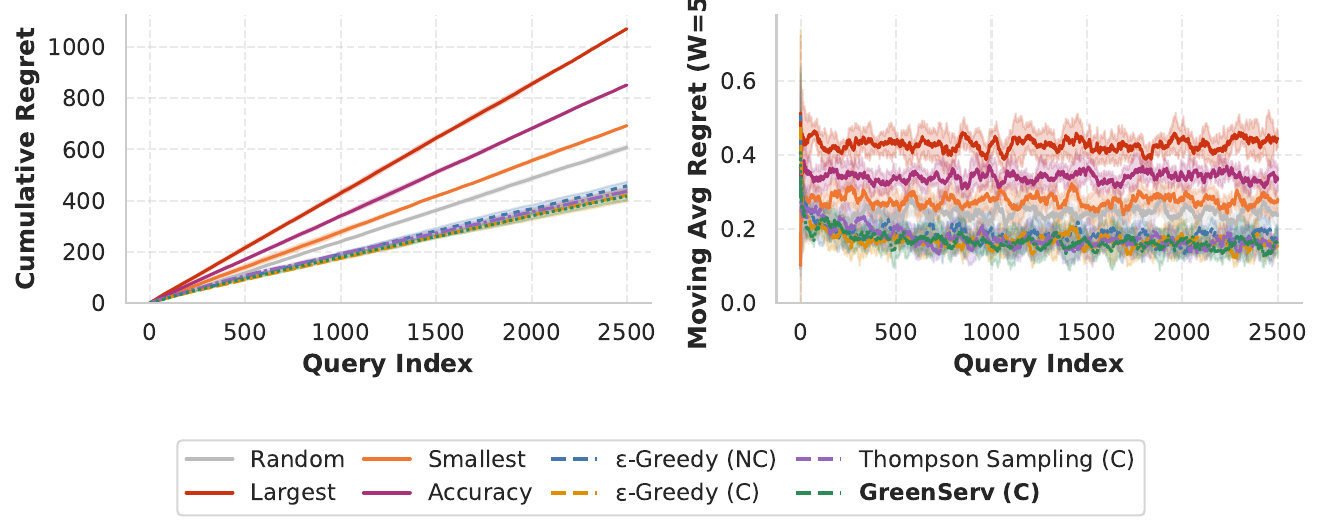}
    \caption{The left plot shows the cumulative regret over time for \approachName and baseline MAB algorithms using the Full Features context. Shaded areas represent 95\% confidence intervals. The right plot shows the moving average  (window=50) regret over time for \approachName and baseline MAB algorithms using the Full Features context, smoothing short-term fluctuations.}
    \label{fig:regret-comparison}
\end{figure}

Figure~\ref{fig:regret-comparison} (left) illustrates the cumulative regret of \approachName and the baseline MAB algorithms over the query sequence. The plot confirms the expected linear regret increase of the static baselines and the random selection. The non-contextual \(\epsilon\)-Greedy implementation demonstrates learning capacity. However, its regret grows visibly faster than \approachName and the contextual baselines. \approachName and the contextual baselines show more effective learning, indicated by their flatter regret curves visually beginning to diverge after the initial 200-300 queries. This, on the other hand, shows the impact of early policy optimization on overall regret. Overall, Contextual \(\epsilon\)-Greedy yields the lowest mean regret (\( \approx 398 \)), around 15\% reduction compared to its non-contextual version (\( \approx 466 \)). \approachName using LinUCB (\( \approx 412 \)) and Contextual Thompson Sampling (\( \approx 400 \)) achieve similar regret reductions.

Figure~\ref{fig:regret-comparison} (right) visualizes the moving average regret for a window size of 50 to investigate learning stability and convergence patterns. An initial \emph{cold start} period is evident for \approachName and the baseline algorithms. We can observe a high and erratic average regret during the first \( \approx 50 \) steps of exploration. Following this phase, both non-contextual and contextual algorithms stabilize relatively quickly, with the non-contextual \(\epsilon\)-Greedy stabilizing at a higher average regret level (\( \approx 0.18 \) vs. \( \approx 0.16 \)). In contrast, \approachName and the contextual baselines converge slower when compared to the non-contextual ones. The frequently crossing lines of \approachName and the contextual baselines visible in Figure~\ref{fig:regret-comparison} (right) suggest that learned policies may be similar yet not identical. We report in Appendix~\ref{appendix:model-selection-patterns} additional results that illustrate these nuances in model selection patterns across \approachName and the baseline algorithms.

\subsubsection{Trade-off Analysis (\(\lambda\) Sweep)}
\label{sec:results_lambda}
\begin{figure}[!ht]
    \centering
    \includegraphics[width=\linewidth]{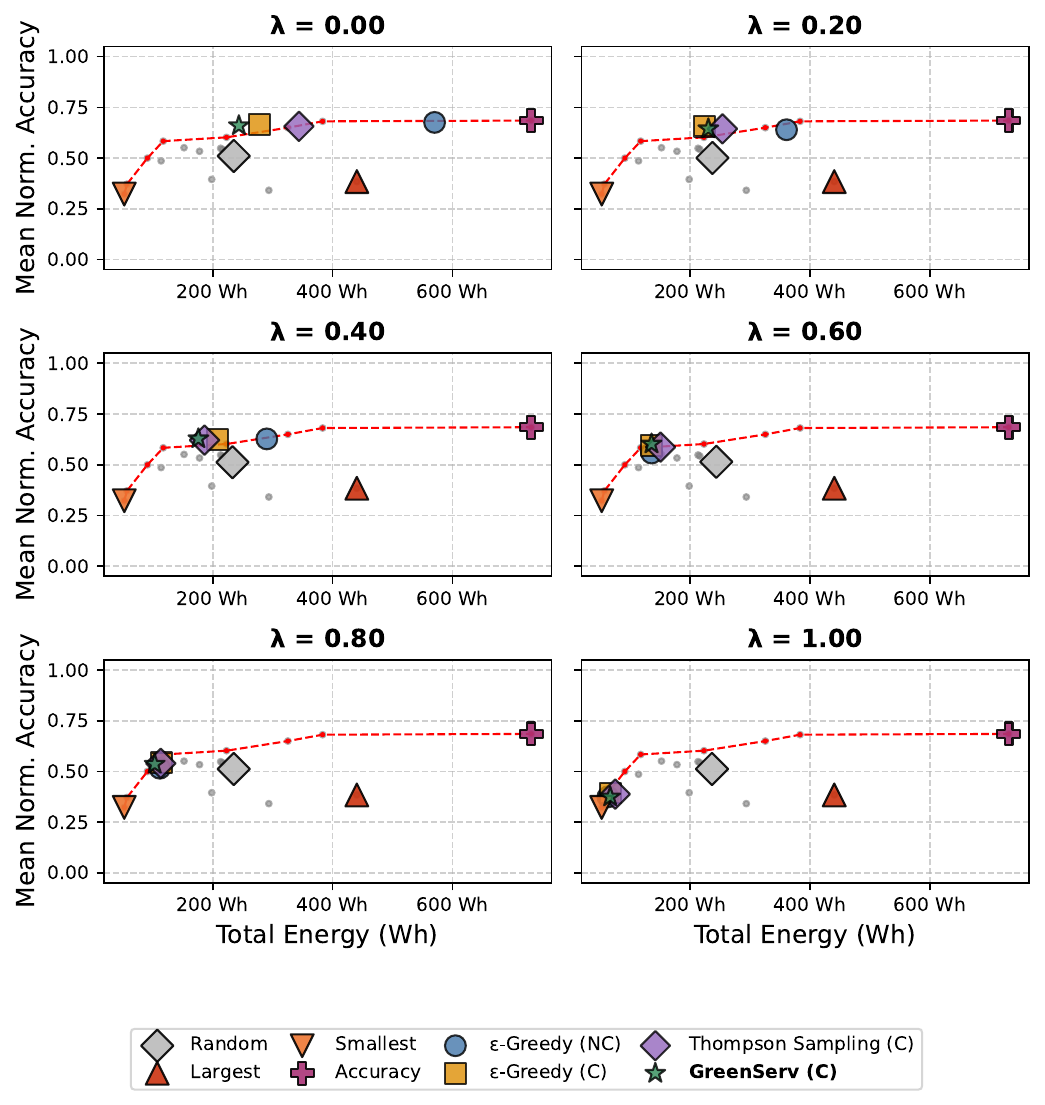}
    \caption{Mean Normalized Accuracy vs. Total Energy Consumption (Wh) for different strategies at varying \(\lambda\) values (0.0 to 1.0, increments of 0.2). The static Pareto front is shown for reference.}
    \label{fig:a4_lambda_pareto}
\end{figure}

Figure \ref{fig:a4_lambda_pareto} illustrates the trade-off between Mean Normalized Accuracy and Total Energy Consumption (Wh) across different \(\lambda\) values. As \(\lambda\) increases, MAB results follow the Pareto front from upper-right to lower-left. Remarkably, \approachName and the contextual baselines consistently operate close to or beyond the static Pareto front (red dashed line). A detailed sensitivity analysis of \(\lambda\), including absolute values and baseline comparisons, is presented in Appendix \ref{sec:results_lambda}.

\subsubsection{Impact of Contextual Features}
\label{sec:results_features}

\begin{figure}[!ht]
    \centering
    \includegraphics[width=0.9\linewidth]{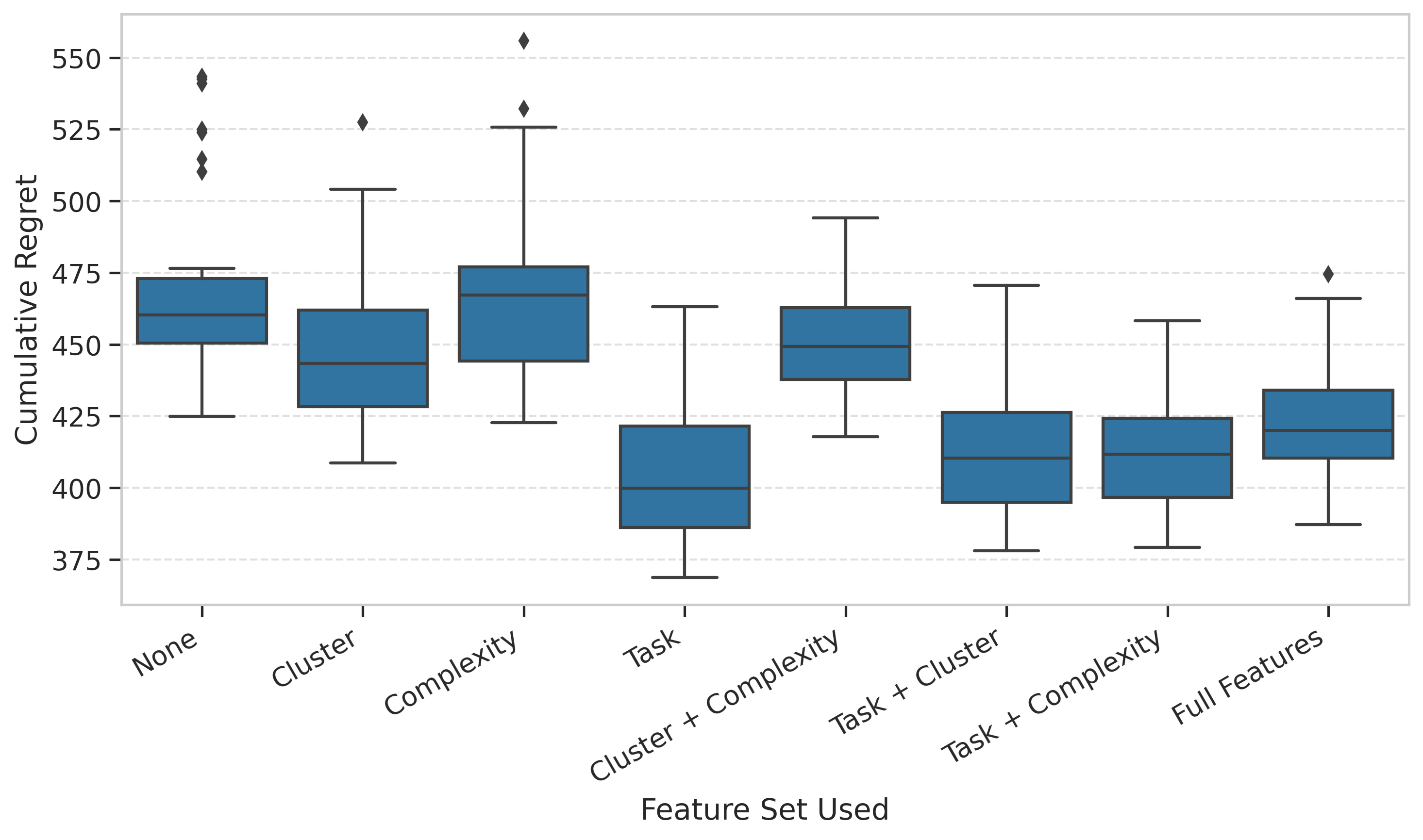} 
    \caption{Distribution of final cumulative regret across multiple runs (n=50) for different contextual feature configurations at \(\lambda=0.4\).}
    \label{fig:a3_feature_boxplot}
\end{figure}

Figure~\ref{fig:a3_feature_boxplot} presents the final cumulative regret distribution across 50 independent runs for each feature configuration: \emph{None} (context-free baseline), \emph{Task}, \emph{Cluster}, \emph{Complexity}, \emph{Task + Cluster}, \emph{Task + Complexity}, \emph{Cluster + Complexity}, and \emph{Full} (all features).

On average, Cluster reduces regret (-17) while Complexity increases it slightly (+7) compared to the implementation without features. However, the most substantial reduction in regret appears to be linked to the inclusion of the \emph{Task} feature, dropping median cumulative regret to \( \approx 400 \). This identifies the task type as the single most informative component of context for guiding model selection in our setup. Combinations involving the \emph{Task} feature (\emph{Task + Cluster}, \emph{Task + Complexity}) retain or slightly improve this level of regret reduction. However, including all features appears to raise regret levels notably. This might be attributed to the increased dimensionality which potentially slows convergence for the MABs during learning or introduce noise from less informative feature interactions compared to the strong signal provided by the task type alone. We report in Appendix~\ref{appendix:contexual-features} additional results that present how context influences model selection behavior by showing the selection frequency of each model.

\subsubsection{Adaptability: Model Addition}
\label{sec:results_adaptability}

\begin{figure}[!ht]
    \centering
    \includegraphics[width=1.05\linewidth]{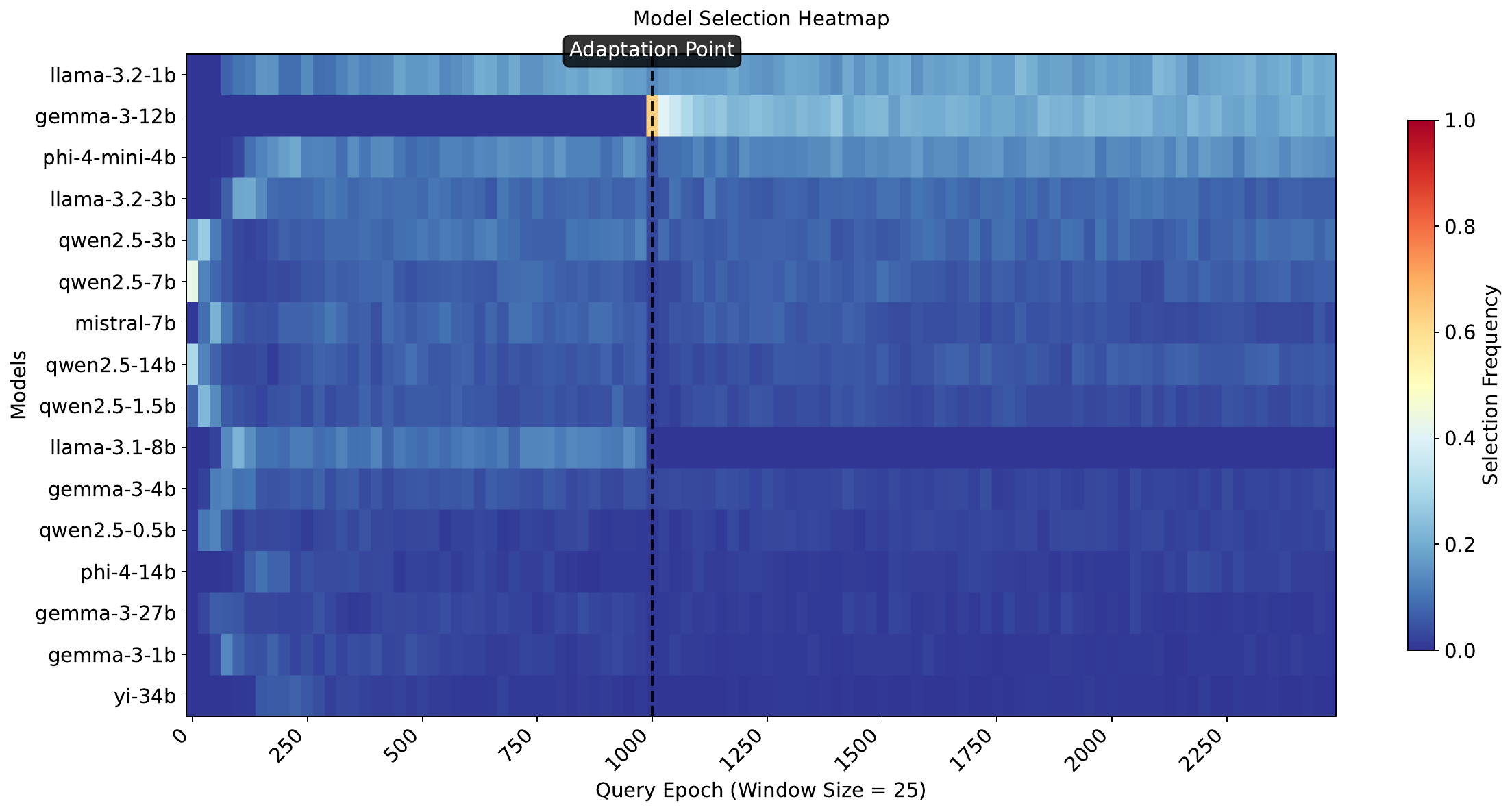}
    \caption{Mean selection frequency over time using \approachName (\(\lambda=0.2\)) for a single run. Gemma-3-12b is added at Query Step 1000 (black line).}
    \label{fig:a6_adaptability_add}
\end{figure}

Figure~\ref{fig:a6_adaptability_add} shows the mean selection frequency of each model over the course of the experiment with a window size of 25. Before the adaptation point (black line), the newly added model (Gemma-3-12b) correctly shows zero selection frequency as it was not yet part of the pool. Immediately after its introduction, the algorithm begins to explore it, and its selection frequency rapidly increases. After around 100 queries, the selection frequency stabilizes at around 20\%-25\%. This shift in selection frequency visibly comes at the cost of previously frequently selected models (LLama-3.1-8b, Gemma-3-27b, Gemma-3-1b), which indicates the successful adaptation of the system to the pool of models by integrating the model into its routing strategy.

\subsubsection{Overhead Analysis}\label{sec:results_overhead}

The average overhead introduced by the system for each query consists of several distinct components. Task type classification requires 3.04~ms, semantic cluster identification takes 3.37~ms, and text complexity calculation adds 0.15~ms. For \approachName, the LinUCB routing decision adds 0.86~ms. Summing these, the total pre-inference overhead per query is approximately 7.77~ms when processed sequentially.

When compared to the actual inference times of the evaluated models, this overhead is minor. For example, median inference latencies range from 36.1~ms for Llama-3.2-1B to 199.7~ms for Gemma-3-27B, with the relative overhead (assuming a 7.8~ms fixed cost) ranging from 21.6\% for the fastest model down to just 3.9\% for the slowest. We report in Appendix~\ref{appendix:overhead-analysis} the detailed results for each of the models.

In summary, results indicate that the routing and feature extraction pipeline does not constitute a significant bottleneck relative to the overall inference time of state-of-the-art LLMs. In fact, the measured overhead is negligible for batch processing scenarios or applications where latency is not critical. However, in latency-sensitive deployments, further optimization of the feature extraction and routing steps may be warranted to minimize their impact.

\subsubsection{External Benchmark Validation.} We further validated our \approachName using RouterBench~\cite{hu2024routerbench} and evaluated on \( \approx \) 36k queries spanning 9 tasks. Table~\ref{tab:routerbench} presents the results across three key metrics. \approachName achieves the best peak and average accuracy at 75.7\% and 71.7\% respectively, which informed our algorithm selection. Contextual $\epsilon$-Greedy achieves the highest AIQ of 0.637, where AIQ is RouterBench's primary metric for capturing the cost-performance trade-off frontier across different willingness-to-pay (WTP) parameters corresponding to \approachName's $\lambda$.

\begin{table}[!ht]
\caption{Performance of contextual routing algorithms on RouterBench. AIQ is averaged across all 9 tasks.}
\centering
\small
\begin{tabular}{@{}lccc@{}}
\toprule
Algorithm & AIQ & Peak Acc. & Avg. Acc. \\
\midrule
\approachName & 0.607 & \textbf{75.7\%} & \textbf{71.7\%} \\
$\epsilon$-Greedy & \textbf{0.637} & 75.4\% & 69.2\% \\
Thompson Sampling & 0.624 & 69.5\% & 66.3\% \\
\bottomrule
\end{tabular}
\label{tab:routerbench}
\end{table}

\textbf{Summary.} Dynamic routing using contextual bandits consistently outperformed static and random baselines. At \(\lambda = 0.4\), \approachName exceeded the Pareto front, achieving accuracy-energy operating points unreachable by single-model deployments. Compared to baselines: Random (+22\% accuracy, -31\% energy), Smallest (+90-100\% accuracy, +400\% energy), and accuracy-optimized (-10-12\% accuracy, -75\% energy). External validation (RouterBench) showed \approachName's superior accuracy. All contextual approaches achieved comparable performance, which confirms that feature engineering, model pool, and reward design are critical factors.

\subsection{Discussion}\label{sec:discussion}

Our study confirms that the accuracy–efficiency trade-off in LLM inference is significantly influenced by both inherent model characteristics (e.g., parameter count, architecture, training) and query-specific properties. As shown in the experimental results (see ~\S ~\ref{sec:results_analysis}), models demonstrate varying levels of accuracy and resource consumption depending on the task type, domain, and complexity of the input query. \approachName addresses these trade-offs by inherently incorporating contextual information (i.e., query features) and adapting routing decisions based on the varying accuracy–efficiency performance across diverse queries and model repositories.

\approachName includes the following limitations.
\textit{First}, MAB algorithms assume stationary rewards and adapt slowly to drifts. Periodic system calibration could address this.
\textit{Second}, our current evaluation focuses on tasks with objective ground truth (EM, ROUGE scores) to enable deterministic accuracy measurement. Many production LLM deployments involve structured tasks (classification, extraction, QA) where ground truth is available. Our framework can be extended to use alternative quality signals such as user feedback or LLM-as-judge evaluations~\cite{chang2024survey} for open-ended generation tasks.
\textit{Third}, generalization across hardware requires latency profiles for various GPU architectures.
\textit{Fourth}, while we evaluated the impact of contextual characteristics, the sensitivity to specific features engineering choices, such as the number of clusters $K$ or the number of complexity bins $N$, could be further explored. 
\textit{Finally}, the empirical results obtained in this study are based on controlled environments for LLM deployments. In contrast, operational conditions should account for factors such as request concurrency, batch processing, queuing delays and runtime model switching.

\section{Conclusions}\label{sec:conclusions}
We present \approachName, a dynamic LLM inference routing framework that employs multi-armed bandits (MABs) to balance accuracy and energy consumption. It extracts a lightweight and multi-feature query context and leverages online MAB algorithms that adapt routing policies using partial feedback, eliminating the need for costly offline calibration. Formalizing routing as contextual multi-objective optimization with direct GPU energy measurements addresses the limitations of prior approaches that rely on proxy cost metrics.

Evaluation demonstrates superior performance over static baselines, achieving 22\% higher accuracy and a 31\% lower energy consumption under optimal configurations. Results validate the framework's 
ability to adapt policies to new models at runtime without requiring expensive offline recalibration.

Future work will extend the framework to support low-level hardware knob configurations, reducing the energy cost of LLM inference and scale our experiments to include larger models and multi-node cluster deployments.

\newpage

\bibliographystyle{ACM-Reference-Format} %
\bibliography{references}

\appendix
\appendix

\section{Appendix: Experiment Details and Results}

\subsection{Model Pool}\label{appendix:model-pool}

\begin{table}[!ht]
    \centering
    \small
    \caption{Model Pool}
    \label{tab:model_pool}
    \resizebox{0.7\linewidth}{!}{%
    \begin{tabular}{llcl}
        \toprule
        \textbf{Family} & \textbf{Version} & \textbf{\# Parameters (B)} & \textbf{HF Handle} \\ 
        \midrule
        \multirow{5}{*}{Qwen} 
            & 2.5 & 0.5 & Qwen/Qwen2.5-0.5B-Instruct \\ 
            & 2.5 & 1.5 & Qwen/Qwen2.5-1.5B-Instruct \\
            & 2.5 & 3   & Qwen/Qwen2.5-3B-Instruct   \\
            & 2.5 & 7   & Qwen/Qwen2.5-7B            \\
            & 2.5 & 14  & Qwen/Qwen2.5-14B-Instruct  \\
        \midrule
        Mistral & v0.3 & 7 & mistralai/Mistral-7B-Instruct-v0.3 \\
        \midrule
        \multirow{4}{*}{Gemma} 
            & 3 & 1  & google/gemma-3-1b-it \\ 
            & 3 & 4  & google/gemma-3-4b-it \\ 
            & 3 & 12 & google/gemma-3-12b-it \\ 
            & 3 & 27 & google/gemma-3-27b-it \\ 
        \midrule
        \multirow{3}{*}{Llama} 
            & 3.1 & 1 & meta-llama/Llama-3.1-1B-Instruct \\ 
            & 3.2 & 3 & meta-llama/Llama-3.2-3B-Instruct \\ 
            & 3.1 & 8 & meta-llama/Llama-3.1-8B-Instruct \\ 
        \midrule
        \multirow{2}{*}{Phi} 
            & 4-mini & 4 & microsoft/Phi-4-mini-instruct \\ 
            & 4      & 14 & microsoft/Phi-4-14B           \\ 
        \midrule
        Yi & - & 34 & 01-ai/Yi-34B \\
        \bottomrule
    \end{tabular}
    } 
\end{table}

Table~\ref{tab:model_pool} lists the LLMs used in the experiments, grouped by model family, along with their parameter counts and Hugging Face identifiers (HF Handles).

\subsection{Model Selection Patterns}\label{appendix:model-selection-patterns}

\begin{figure}[!ht]
    \centering
    \includegraphics[width=\linewidth]{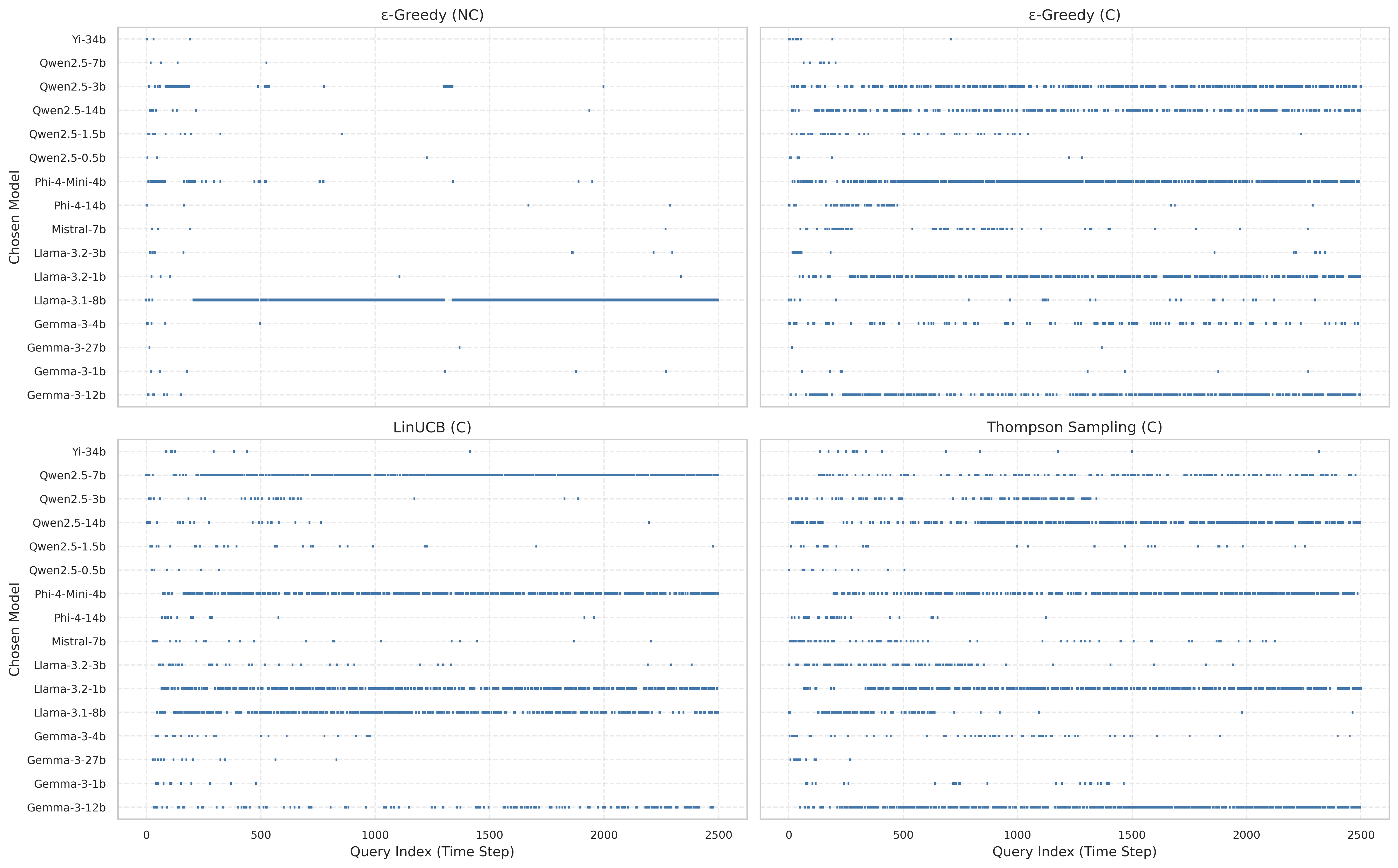}
    \caption{Sequence of models chosen by the MAB algorithms during a single run (\(\lambda=0.4\)).}
    \label{fig:a5_selection_timeline}
\end{figure}

Figure \ref{fig:a5_selection_timeline} illustrates model selection patterns across the MAB algorithms. Models like Llama-3.1-8B and Phi-4-Mini-4B show high selection frequency across algorithms. Contextual algorithms (C) exhibit more distributed patterns than non-contextual (NC) variants, indicating finer-grained performance distinctions, particularly evident in middle-tier models. This suggests algorithms identify niches where certain models excel despite not being globally optimal.

\subsection{Contextual Features}\label{appendix:contexual-features}

\begin{figure}[!ht]
    \centering
    \begin{minipage}{\linewidth}
        \centering
        \includegraphics[width=\textwidth]{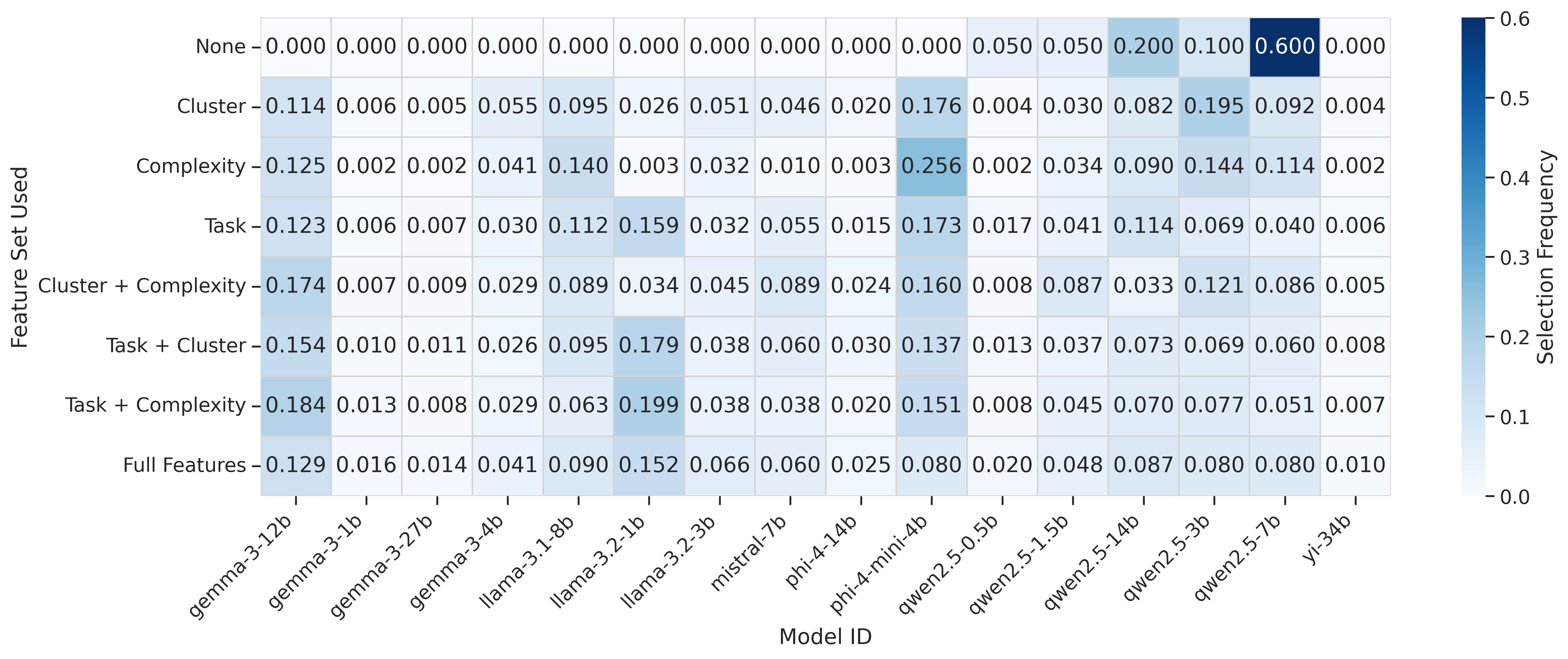}
    \end{minipage}
    
    \vspace{0.1cm}
    
    \begin{minipage}{\linewidth}
        \centering
        \includegraphics[width=\textwidth]{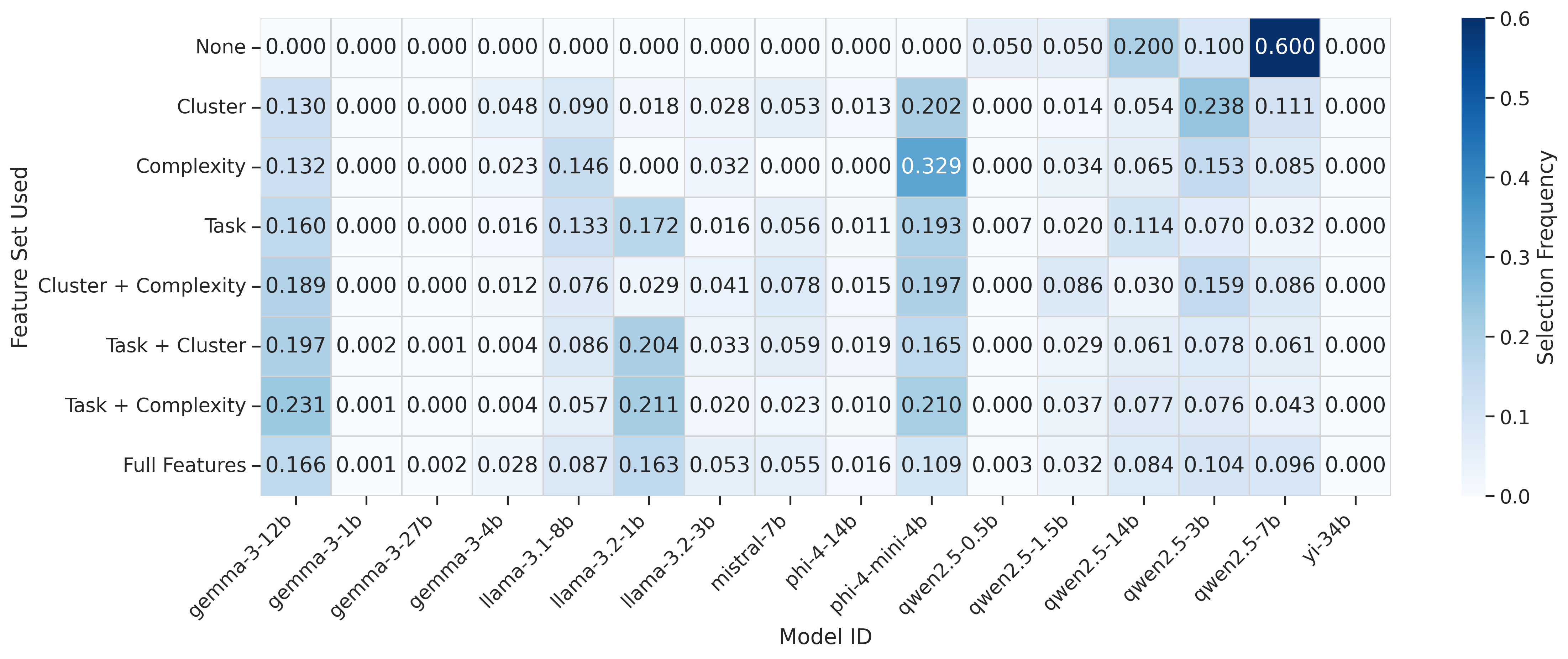}
    \end{minipage}
    
    \caption{The heatmaps show the selection frequency of each model for different feature configurations across runs (n=10). The top heatmap shows selections during the first half of the experiments (1-1250), while the bottom shows the second half (1251-2500). Darker blue indicates higher selection frequency. Without contextual features (None), selections concentrate on fewer models, while adding features leads to more diverse selection patterns that become increasingly focused as the algorithm learns.}
    \label{fig:feature_selection_heatmaps}
\end{figure}

Figure~\ref{fig:feature_selection_heatmaps} shows how context influences model selection frequency. The top heatmap displays patterns for the first half (1-1250), while the bottom shows the second half (T=1251 to T=2500), averaged over ten runs. Initially, selections distribute widely as algorithms explore. Adding features increases exploration, further distributing loads across models. In contrast, the second half shows more concentrated selection patterns as policies stabilize. The baseline without features consistently favors Qwen2.5-7B (\( \approx 0.6 \) frequency in the second half), while contextual approaches develop more sophisticated strategies. When the \emph{task} feature is included, the Contextual MAB appears to prefer Llama-3.2-1B (\( \approx 0.17 \)) and Phi-4-mini-4B (\( \approx 0.19 \)) among others. The \emph{Full Features} configuration demonstrates the most spread-out policies as it tries to match query requirements precisely.

\subsection{Trade-off Analysis (\(\lambda\) Sweep)}
\label{sec:results_lambda}

\begin{figure}[!ht]
    \centering
    \includegraphics[width=\linewidth]{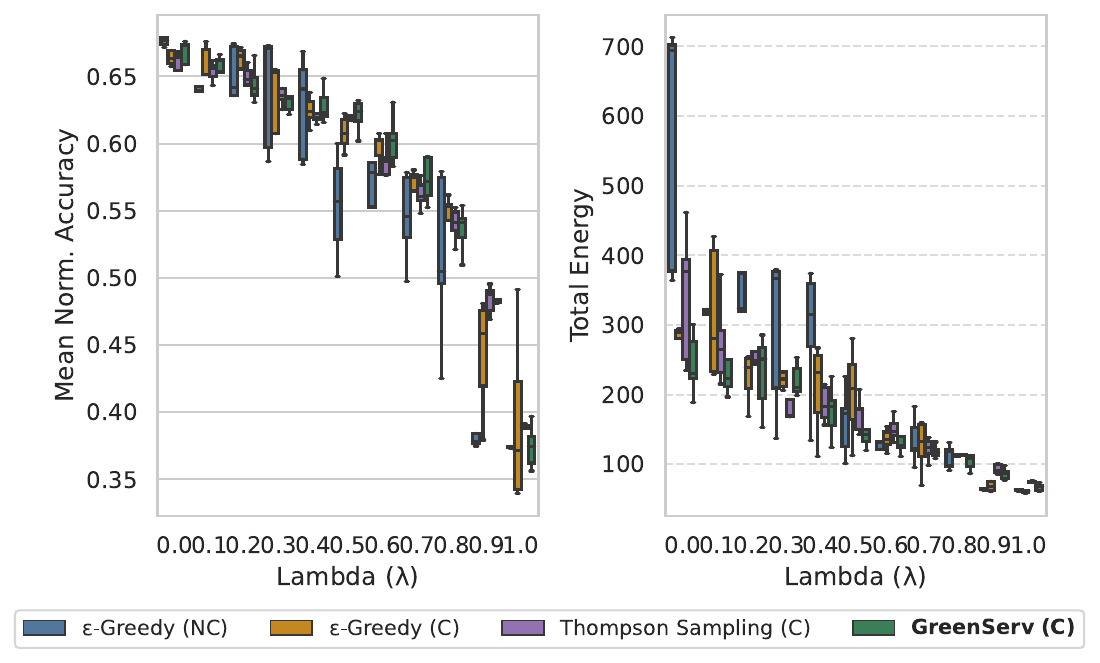} 
    \caption{Distribution of Mean Normalized Accuracy (top) and Total Energy Consumption (bottom) for \approachName and baseline MAB strategies across \(\lambda\) values.}
    \label{fig:a4_lambda_boxplot}
\end{figure}

Figure~\ref{fig:a4_lambda_boxplot} presents the distribution of mean normalized accuracy and total energy consumption across 20 runs for \approachName and baseline algorithms as \(\lambda\) changes in between 0 and 1. %
Both accuracy and energy consumption decrease as \(\lambda\) increases, which demonstrates the system's ability to prioritize either objective when instructed. \approachName and the contextual baselines show similar trends, maintaining slightly higher accuracy, lower energy consumption and greater robustness compared to the non-contextual \(\epsilon\)-Greedy across most \(\lambda\) values.

Figure~\ref{fig:a4_lambda_pareto} provides another perspective on the trade-off by showing our strategy aggregates on accuracy-energy for different \(\lambda\) values (in increments of 0.2 for clarity). Each point represents the average accuracy-energy outcome of an algorithm at a specific \(\lambda\) value. As \(\lambda\) increases, MAB results follow the Pareto front from upper-right to lower-left. Remarkably, \approachName and the contextual baselines consistently operate close to or beyond the static Pareto front (red dashed line).

\subsection{Overhead Analysis}\label{appendix:overhead-analysis}

\begin{table}[!ht]
    \centering
    \caption{Model Inference Latency Statistics}
    \label{tab:model_latency_stats}
    \resizebox{0.9\linewidth}{!}{%
    \begin{tabular}{lrrrrrr|r}
        \toprule
        \textbf{Model} & \textbf{Min} & \textbf{P25} & \textbf{Median} & \textbf{Average} & \textbf{P75} & \textbf{Max} & \textbf{Overhead (\%)} \\
        \midrule
        Llama-3.2-1B & 19.8 & 34.6 & 36.1 & 570.8 & 1,196.7 & 2,401.6 & 21.6 \\
        Qwen2.5-0.5B & 14.0 & 43.2 & 44.3 & 686.1 & 1,402.9 & 3,350.8 & 17.6 \\
        Qwen2.5-1.5B & 20.0 & 48.9 & 50.6 & 976.1 & 1,926.0 & 3,907.8 & 15.4 \\
        Llama-3.2-3B & 16.4 & 47.1 & 51.6 & 977.9 & 1,955.6 & 3,936.4 & 15.1 \\
        Phi-4-mini-4B & 18.3 & 47.6 & 52.7 & 982.6 & 1,605.1 & 4,557.5 & 14.8 \\
        Qwen2.5-3B & 40.2 & 50.4 & 53.8 & 1,351.4 & 2,078.4 & 4,995.9 & 14.5 \\
        Qwen2.5-7B & 38.4 & 48.0 & 56.4 & 1,880.2 & 3,525.7 & 10,523.9 & 13.8 \\
        Llama-3.1-8B & 42.1 & 51.5 & 60.8 & 1,366.5 & 2,491.6 & 4,603.7 & 12.8 \\
        Mistral-7B & 42.1 & 51.4 & 60.9 & 1,103.2 & 2,345.4 & 4,324.3 & 12.8 \\
        Gemma-3-1B & 52.4 & 57.3 & 68.2 & 2,848.3 & 5,026.1 & 12,666.0 & 11.4 \\
        Phi-4-14B & 33.9 & 62.7 & 76.8 & 1,553.8 & 2,420.8 & 6,790.7 & 10.2 \\
        Gemma-3-4B & 72.6 & 75.8 & 81.1 & 2,225.8 & 3,595.8 & 9,039.5 & 9.6 \\
        Gemma-3-12B & 48.2 & 61.3 & 83.0 & 3,104.4 & 5,771.0 & 14,296.9 & 9.4 \\
        Qwen2.5-14B & 66.6 & 70.7 & 83.8 & 1,786.2 & 2,208.0 & 7,387.5 & 9.3 \\
        Yi-34B & 71.5 & 113.8 & 163.4 & 2,195.5 & 3,924.0 & 14,571.3 & 4.8 \\
        Gemma-3-27B & 158.4 & 167.1 & 199.7 & 5,088.3 & 6,639.8 & 37,972.6 & 3.9 \\
        \bottomrule
    \end{tabular}
    }
\end{table}

Table~\ref{tab:overhead_components} lists the average elapsed time (ms) for each step involved in the feature extraction and routing decision process for a single query. Combined, the total average overhead per query added by our system is approximately 6.68-7.77~ms when processed sequentially. This overhead should be evaluated in perspective to the actual inference times, which varied significantly across our model pool, as detailed in Table~\ref{tab:model_latency_stats}.

\begin{table}[!ht]
    \centering
    \caption{Average Overhead per Component}
    \label{tab:overhead_components}
    \resizebox{0.9\linewidth}{!}{%
    \begin{tabular}{lr}
        \toprule
        \textbf{Component} & \textbf{Avg. Time per Query (ms)} \\
        \midrule
        Task Type Classification & 3.04 \\
        Semantic Cluster Identification & 3.37 \\
        Text Complexity Calculation & 0.15 \\
        \(\epsilon\)-Greedy Routing Decision & 0.02 \\
        LinUCB Routing Decision & 0.86 \\
        Contextual Thompson Sampling Routing Decision & 1.21 \\
        \midrule
        \textbf{Total Pre-Inference Overhead} & \textbf{6.68-7.77} \\ 
        \bottomrule
    \end{tabular}
    }
\end{table}

\section{Appendix: Complexity Analysis}\label{appendix:complexity-analysis}
The computational complexity of \approachName stems mainly from feature extraction and model selection during routing. We analyze how the system scales with key parameters: the number of queries $T$, model pool size $|M|$, context vector dimension $d$, and average query length $l$.

\subsection{Time Complexity}
For each incoming query, we perform feature extraction followed by routing. In the demonstrated implementation, feature extraction involves computing two transformer embeddings using all-MiniLM-L6-v2, which has a fixed maximum sequence length of 256 tokens. Any input exceeding this limit is truncated, bounding the self-attention computation to $O(256^2) = O(1)$ time regardless of query length. The Flesch Reading Ease calculation adds an $O(l)$ pass through the full text. Since the remaining operations (task classification, cluster assignment and one-hot encoding) require constant time, feature extraction totals $O(l)$ per query, though in practice the constant-time transformer operations dominate.

The routing complexity depends on the chosen algorithm. All variants first check feasibility constraints for each model in $O(|M|)$ time. Non-contextual $\epsilon$-Greedy then requires at most $O(|M|)$ comparisons to find the best model. Contextual algorithms cause higher costs: contextual $\epsilon$-Greedy computes $|M|$ dot products of dimension $d$, resulting in $O(|M|d)$ complexity. LinUCB and Thompson Sampling must invert $d \times d$ matrices for each model, resulting in $O(|M|d^3)$ complexity which is dominated by the matrix operations.
Processing all $T$ queries sequentially results in a time complexity $O(T \cdot (l + |M|d^3))$. With our experimental parameters ($|M|=16$, $d=12$), constant-time transformer embedding dominates feature extraction at approximately 6-7 milliseconds per query, and routing adds 0.02-1.21 milliseconds depending on the algorithm.

\subsection{Space Complexity}
Space complexity remains independent of $T$ as the system maintains only derived statistics. Feature extraction requires $O(K \cdot d{\text{emb}})$ for semantic cluster centroids and $O(n{\text{tasks}} \cdot d_{\text{emb}})$ for classifier weights. MAB algorithms vary in memory usage. Non-contextual $\epsilon$-Greedy requires $O(|M|)$ and its contextual variants $O(|M| \cdot d)$, while LinUCB and Thompson Sampling require $O(|M| \cdot d^2)$ to store the matrices and vectors per model. With $|M|=16$, $d=12$, and $K=3$, total memory usage remains negligible compared to employed language model weights.

\subsection{Practical Implications}
Feature extraction is dominated by constant-time transformer embeddings, while Flesch complexity scales linearly but contributes minimally. For routing, the number of models $|M|$ and context vector dimension $d$ affect execution time. The cubic scaling in $d$ for contextual bandits appears concerning but remains manageable with $d=12$ when applying modern libraries that optimize these matrix operations. Combined, our framework achieves linear time scaling with respect to the primary input size $T$ while maintaining constant space complexity. The combination of predictable per-query cost and fixed memory usage makes the system suitable for long-running deployments processing millions of queries.

\end{document}